\documentclass{nature}

\usepackage{epsfig}
\usepackage{color}
\usepackage{amsmath}
\usepackage{lscape}
\usepackage{cite}
\usepackage{amssymb}

\begin{document}
\title{Carbon-dioxide-like Skyrmion controlled by spin-orbit coupling in atomic-molecular Bose-Einstein condensates}

\author{Chao-Fei Liu$^{1,2}$, Gediminas Juzeli\={u}nas$^{3}$, and Wu-Ming Liu$^{1\star}$}
\maketitle

\begin{affiliations}
\item
Beijing National Laboratory for Condensed Matter Physics,
Institute of Physics, Chinese Academy of Sciences,
Beijing 100190, China

\item
School of Science, Jiangxi University of Science and Technology, Ganzhou 341000, China


\item
Institute of Theoretical Physics and Astronomy, Vilnius University, A. Go$\breve{s}$tauto 12, 01108 Vilnius, Lithuania


$^\star$e-mail: wliu@iphy.ac.cn

\end{affiliations}

\begin{abstract}

Atomic-molecular Bose-Einstein condensates (BECs) offer brand new opportunities to revolutionize quantum gases
and probe the variation of fundamental constants with unprecedented sensitivity.
The recent realization of spin-orbit coupling (SOC) in BECs provides a new platform
for exploring completely new phenomena unrealizable elsewhere.
However, there is no study of SOC atomic-molecular BECs so far.
Here, we find a novel way of creating a Rashba-Dresselhaus SOC in atomic-molecular BECs by combining the spin dependent photoassociation and Raman coupling, which can control the formation and distribution of a new type of topological excitation
-- carbon-dioxide-like Skyrmion.
This Skyrmion is formed by two half-Skyrmions of molecular BECs coupling with one Skyrmion of atomic BECs,
where the two half-Skyrmions locates at both sides of one Skyrmion,
which can be detected by measuring the vortices structures using the time-of-flight absorption imaging technique in real experiments.

\end{abstract}

The production of atomic-molecular Bose-Einstein condensates (BECs) via the Raman photoassociation \cite{ Donley, Heinzen, Wynar, McKenzie, YanMi, Chenyong1, Chenyong2} has revolutionized the field of ultracold quantum gases and generated much interest in controlled molecular dynamics, ultrahigh-resolution molecular spectroscopy and quantum information processing, where molecules offer microscopic degrees of freedom beyond atomic gases. Atomic-molecular BECs possess unique properties that may allow for the study of new physical phenomena and lead to discoveries, reaching far beyond the atomic BEC.
Furthermore, atomic-molecular BECs exhibit the non-Arrhenius superchemistry, which describes the coherent stimulation of chemical reactions via macroscopic occupation of a quantum state by a bosonic chemical species \cite{Jinghui, Donley, Heinzen, Wynar, McKenzie, Drummo, Hope, YanMi, Chenyong1, Chenyong2}.
Recently, spin-orbit coupling (SOC) has been realized for ultracold gases of atomic bosons \cite{Lind, Shuai-Chen2, Williams, Hamner, ChenYong, Zhangchuanwei} and fermions \cite{CheukLW, ZhangjingFermi}. Especially, SOC is ubiquitous in physical systems.
This raises a question how to create SOC in the mixed atomic-molecular BECs. At least, the hyperfine states of ultracold molecules \cite{ StraussC, Aldegunde, Hamley} demonstrate that molecular systems possess the essential characteristics for the creation of SOC.
However, there is no study of the SOC atomic-molecular BECs so far.

SOC describes the interaction of the particle's spin with its orbital motion.
It represents a fascinating and fast-developing area of research significantly overlapping with traditional condensed matter physics, but importantly containing also completely new phenomena unrealizable elsewhere.
Novel effects have been reported, such as formation of the Bose mixture exhibiting a phase-separated propagation \cite{Lind},
as well as appearance of the plane wave and stripe phases \cite{Wang, TinLun, LiYun}
and other nontrivial structures \cite{Demler, Sinha, xuxiaoqiang, Huhui, Wucj2, Wilson, Achilleos, Kawakami, Lobanov, Konotop, Malomed1, Malomed2} in the SOC atomic BECs.
A number of schemes have been proposed to create general gauge fields and SOC for ultracold atoms
\cite{Ruseckas, Stanescu, Juzeliunas2010, SOC_Anderson, SOC_Anderson2}.
Further exploration of SOC on atomic-molecular BECs is, however, still lacking.
Differing from the normal SOC ultracold atoms, the SOC atomic-molecular BECs would initiate the exploration of SOC to the untouched hybrid quantum systems.

Here, we find a novel way of creating a Rashba-Dresselhaus SOC in the atomic-molecular BECs by combining the spin dependent photoassociation \cite{Hamley, Kobayashi} and Raman coupling \cite{Lind, Shuai-Chen2, Williams, Hamner, ChenYong, Zhangchuanwei}.
By applying this SOC, we demonstrate that the atomic-molecular BECs support a new type of topological spin texture -- carbon-dioxide-like Skyrmion, which is formed by two half-Skyrmions of the molecular BECs coupling one Skyrmion of the atomic BECs, where the former two half-Skyrmions locates at both sides of the Skyrmion.
Our results not only indicate the immanent mechanism of a single carbon-dioxide-like Skyrmion, but also show that SOC can control the carbon-dioxide-like Skyrmions to distribute in parallel along the $x$ axis.
The SOC even can destroy the half-Skyrmion in the carbon-dioxide-like Skyrmions which are in parallel along the $x$ axis when its strength is very strong.
In addition, our results prove that the atom-molecule conversion, which describes the superchemistry, causes the combination of the Skyrmion and half-Skyrmions in this carbon-dioxide-like Skyrmion.
If there is no the SOC or the accompanying Raman coupling, the carbon-dioxide-like Skyrmion will not occur.
Furthermore, we find that the formation of the carbon-dioxide-like Skyrmion relates to some vortices structures. For instance, the Skyrmion in the carbon-dioxide-like Skyrmion may originate from a dipole structure of a pair of vortices in the atomic BECs, and the half-Skyrmion in the carbon-dioxide-like Skyrmion may derive from a three-vortex structure, where two vortices of $m_{F}=-1$ and $m_{F}=1$ components are respectively on both sides of a vortex of $m_{F}=0$ component in spin-1 molecular BECs.
Therefore, the detection of carbon-dioxide-like Skyrmion can be based on the vortices structures by using the absorption imaging technique in a real experiment.

\section*{Results}

\subsection{Creation of Rashba-Dresselhaus spin-orbit coupling in atomic-molecular BECs.}

Experimentally, the Rashba-Dresselhaus SOC can be obtained by using the Raman-coupling of two hyperfine states in the ground state manifold of alkali atoms \cite{Lind, Shuai-Chen2, Williams, Hamner, ChenYong, Zhangchuanwei}. In that case, the strength of the SOC depends on the wave vectors of the Raman lasers. On the other hand, the effective
out-of-plane (in-plane)
Zeeman field is proportional to the Rabi frequency of the Raman process, i.e.
to the matrix element of the two-photon coupling between a selected pair of atomic magnetic sublevels of the ground-state manifold.
It is well known that the molecular BEC can be created by the Raman photoassociation of atoms in a condensate \cite{ Donley, Heinzen, Wynar, McKenzie, YanMi, Chenyong1, Chenyong2}. With the spin dependent photoassociation \cite{Hamley, Kobayashi},
the hyperfine states of molecules are related to those of the converted atoms.
Furthermore, a number of experiments have demonstrated a high degree of control over the hyperfine state of ultracold molecules via two-photon optical Raman transitions \cite{Hamley, Kobayashi, Yejun, Yejun2}.
Here, when considering the SOC in atomic-molecular BECs, we are dealing with the atoms characterized by two different internal (quasi-spin) states,
as well as molecules with three spin states ($F=1, m_{F}=0, \pm1$).

We consider a Bose gas of ultracold $^{87}$Rb atoms, commonly used in both experiments on the SOC atomic
BEC \cite{Lind, Shuai-Chen2, Williams, Hamner, ChenYong, Zhangchuanwei} and the atomic-molecular BECs \cite{Heinzen, Wynar}, to demonstrate the scheme for creating SOC atomic-molecular BECs [see Fig. 1a].
We take a BEC of up to $3\times10^{5}$ atoms, and
select two atomic internal spin states with $m_{f}=-1$ and $m_{f}=0$ of the $^{87}$Rb 5S1/2 ground state $f=1$ manifold, to be referred to as the (pseudo-) spin up ($|\uparrow\rangle$) and down ($|\downarrow\rangle$) states [see Fig. 1b]. Here, $f$ and $m_{f}$ are the quantum numbers of the atomic total spin and its projection along a quantisation axis.
A pair of counter-propagating Raman lasers ($\omega_{L3}$ and $\omega_{L4}$)
couple these spin states with the matrix element (Raman Rabi frequency) $\Omega_{A}$ inducing the SOC of the atomic BECs [see Fig. 1b].
The third spin state with $m_{f}=1$ is detuned away from
the Raman resonance due to the quadratic Zeeman shift and hence can be neglected.
On the other hand, the counter-propagating Raman lasers ($\omega_{L5}$, $\omega_{L6}$ and $\omega_{L7}$)
are used to induce the SOC for the spin-1 ($F=1$) molecular BECs [see Fig. 1c]. An extra laser $\omega_{L7}$ is needed to couple the molecular states with $m_F=0$ and $m_F=1$ detuned from the resonance between the states $m_F=-1$ and $m_F=0$ due to a quadratic Zeeman effect.
To prevent population of the excited state the laser fields
are to be sufficiently detuned from the single photon resonance. The detailed processes for characterizing the SOC atomic-molecular BECs are indicated in the Methods.

The molecular BECs of $^{87}$Rb$_{2}$ with hyperfine state $|F=1$, $m_{F}=0,\pm1\rangle$ are produced by the spin dependent photoassociation of atoms \cite{Hamley, Kobayashi} (see Figs. 1d-1f).
A pair of atoms with various combinations of internal states
is coupled to an excited molecule state $|\phi_{em}\rangle$
via a laser beam $\omega_{L1}$.
Here, the photoassociation through the total spin-2 scattering channel gives molecular $F$ numbers of 2.
The three spin dependent photoassociation process can be written as:
$|1,0\rangle\otimes|1,0\rangle\longrightarrow|2,0\rangle$, $|1,-1\rangle\otimes|1,0\rangle\longrightarrow|2,-1\rangle$ and $|1,-1\rangle\otimes|1,-1\rangle\longrightarrow|2,-2\rangle$.
Under the transition $\Delta F=-1$ and $\Delta m_{F}=+1$, the excited molecule states further change into $|\phi_{em}^{'}\rangle$ with $F=1$ and $I=3$ \cite{Hamley}, where $F$, $m_{F}$ and $I$ denote the total spin, its projection and nuclear spin  quantum numbers of a molecule.
Finally, $|\phi_{em}^{'}\rangle$ is coupled to
a ground molecule state $|\phi_{m}\rangle$ via another laser beam $\omega_{L2}$ [see Figs. 1d-1f].
As a result, the two-component atomic BEC can be coherently converted to the
three-component (spin-1) molecular BEC with three internal spin
states $|F=1,\,m_{F}=0\,,\pm1\rangle$.

The SOC atomic-molecular BECs can be described by (see Methods):
\begin{eqnarray}\label{GP1}
i\hbar\frac{\partial\psi_{a}}{\partial t}&=&[-\frac{\hbar^{2}\nabla^{2}_{d}}{2M_{A}}+\frac{M_{A}\omega^{2}r^{2}}{2}+\sum_{q=\uparrow,\downarrow}g_{a,q}|\psi_{q}|^{2}+\sum_{m=0,\pm1}g_{a,m}|\phi_{m}|^{2}]\psi_{a} \notag \\
&+&\lambda\sum_{q=\uparrow,\downarrow}(\widehat{\sigma}_{z})_{aq}p_{x}\psi_{q}+\hbar\Omega_{A}\sum_{q=\uparrow,\downarrow}(\widehat{\sigma}_{x})_{aq}\psi_{q}+K_{a},
\end{eqnarray}
\begin{eqnarray}\label{GP2}
i\hbar\frac{\partial\phi_{m}}{\partial t}&=&[-\frac{\hbar^{2}\nabla^{2}_{d}}{2M_{M}}+\frac{M_{M}\omega^{2}r^{2}}{2}+g_{n}|\mathbf{\phi}|^{2}+\sum_{q=\uparrow,\downarrow}g_{m,q}|\psi_{q}|^{2}]\phi_{m} \notag \\
&+&g_{s}\sum_{\alpha=x,y,z}\sum_{n,k,l=0\pm1}(\widehat{F}_{\alpha})_{mn}(\widehat{F}_{\alpha})_{kl}\phi_{n}\phi^{*}_{k}\phi_{l} \notag \\
&+&\lambda\sum_{n=0\pm1}(\widehat{F}_{z})_{mn}p_{x}\phi_{n}+2\hbar\Omega_{A}\sum_{n=0\pm1}(\widehat{F}_{x})_{mn}\phi_{n}+K_{m},
\end{eqnarray}
with $K_{\uparrow}=\sqrt{2}\chi\psi_{\uparrow}^{*}\phi_{+1}+\chi\psi_{\downarrow}^{*}\phi_{0}$,  $K_{\downarrow}=\sqrt{2}\chi\psi_{\downarrow}^{*}\phi_{-1}+\chi\psi_{\uparrow}^{*}\phi_{0}$, $K_{+1}=\frac{\chi}{\sqrt{2}}\psi_{\uparrow}^{2}+(\varepsilon+M_{A}\lambda^{2})\phi_{+1}$, $K_{0}=\chi\psi_{\uparrow}\psi_{\downarrow}+\varepsilon\phi_{0}$, and $K_{-1}=\frac{\chi}{\sqrt{2}}\psi_{\downarrow}^{2}+(\varepsilon+M_{A}\lambda^{2})\phi_{-1}$.
Here $a=\uparrow, \downarrow$ and $m=0, \pm1$ labels, respectively, the atomic and molecular spin, $d=1,2,3$ refers to the dimension of the system,
$M_{A}$ ($M_{M}$) is the mass of the atom (molecule), $\widehat{F}_{\alpha=x, y, z}$ ($\widehat{\sigma}_{\alpha=x, y, z}$) are the spin-1 (Pauli) matrices, $g_{a,q}=\frac{4\pi\hbar^{2}a_{aq}}{M_{A}}$ and $g_{a,m}=\frac{2\pi\hbar^{2}a_{am}}{M_{A}M_{M}/(M_{A}+M_{M})}$ are, respectively, the strengths of atom-atom and atom-molecule interaction, $g_{n}$ and $g_{s}$ are the strengths of the spin-independent and spin-dependent interactions between the molecules \cite{HoTLspinor, Ohmispinor}. Here also
$\psi_{a}$ and $\phi_{m}$ denote the macroscopic wave functions of the atomic BEC in the internal state $|a\rangle$ ($a=\uparrow, \downarrow$)
and those of the molecular BEC in the internal state $|m_{F}=m\rangle$ ($m=0, \pm1$).
The parameters $\lambda$, $\Omega_{A}$, $\chi$, $\varepsilon$, and $\omega$ are the strengths of SOC, Raman coupling, atom-molecule conversion, Raman detuning, and trap frequency, respectively.

\subsection{A carbon-dioxide-like Skyrmion in the SOC atomic-molecular BECs.}
Spin texture excitations, such as Skyrmion and half-Skyrmion (meron), are particle-like topological entities in continuous field and are recognized to play an important role in many condensed matter systems \cite{Schmeller, Romming, Nagaosa, XZYu, liuwv1, liuwv2}.
The SOC atomic-molecular BECs provide a new platform for exploring the spin texture excitations coming from atomic BECs \cite{Niuqian1, Niuqian2, Buljan} and molecular BECs simultaneously.
Now, we firstly present a way to obtain a single carbon-dioxide-like Skyrmion, which consists of one Skyrmion with a topological charge $-1$ in the atomic BECs and two half-Skyrmions with a topological charge $-\frac{1}{2}$ in the molecular BECs.
On both sides of the Skyrmion are the two half-Skyrmions and they couple as a unity resembling the carbon dioxide molecule.

In numerical simulations, we set the initial chemical potential $\mu_{m,0}=2\mu_{a,0}=8\hbar\omega$. The final chemical potential of the noncondensate band is increased to $\mu_{m}=2\mu_{a}=28\hbar\omega$.
We use the parameters of atomic-molecular BECs of $^{87}$Rb
with $M_{M}=2M_{A}$ ($M_{A}=144.42\times 10^{-27}Kg$), the trapping frequency $\omega=200\times2\pi$Hz, and strength of atom-atom interaction $g_{\uparrow,\uparrow}=g_{A}$ with the scattering length $a_{A}=101.8a_{B}$, where $a_{B}$ is the Bohr radius.
In addition, when the change in energy of converting two atoms into one molecule ($\Delta U=2U_{Ta}-U_{Tm}$) \cite{Wynar}, without including internal energy, approaches zero, we can obtain the value $2g_{A}=g_{M}$, and we set $g_{n}=g_{M}=2g_{A}$.
For simplify, we set the atom-molecule interaction
$g_{\uparrow,-1}=g_{\uparrow,0}=g_{\uparrow,+1}=g_{\downarrow,-1}=g_{\downarrow,0}=g_{\downarrow,+1}=g_{AM}$.
The unit of length, time, and energy correspond to $\sqrt{\hbar/(M_{A}\omega)}$ ($\approx0.76\mu m$), $\omega^{-1}$ ($\approx0.96 \times 10^{-3}s$), and $\hbar\omega$, respectively.
In the rotating frame, $H$ is replaced by
$H-\Omega_{rotation}\widehat{L}_{z}$, where $\widehat{L}_{z}=-i\hbar(x\partial_{y}-y\partial_{x})$ is the $z$ component of the
orbital angular momentum and $\Omega_{rotation}$ is the angular frequency
of rotation.

The densities and the corresponding phases of the SOC atomic-molecular BECs are obtained for the equilibrium state (see Figs. 2a and 2b). One can see that both components of atomic BECs contain a vortex excitation. The spin-up (spin-down) component fills in the vortex of the spin-down (spin-up) component. Thus, this structure is similar to a vortices molecule \cite{Kasamatsu}.
At the same time, two vortices are induced in each molecular component, respectively.
The densities of all the components are symmetric with respect to the $y$ axis.
In addition, only the density of the $m_{F}=0$ component is symmetric with respect to the $x$ axis.
In this way, these vortices form a cluster.

The emergence of the vortices cluster results from a combination of the SOC, Raman coupling, atom-molecule conversion and rotation effect.
The Rashba-Dresselhaus SOC causes a translational displacement along the $y$ axis on the rotating BECs. The translational displacement for different spin state is opposite. Thus, the spin-up and $m_{F}=+1$ components trend to move to the $y<0$ region, and the spin-down and $m_{F}=-1$ components go to the $y>0$ region. Note that the $m_{F}=0$ component is not affected by the Rashba-Dresselhaus SOC.
Rotation also induces vortices. In the atomic BECs, the spin-up and the spin-down component have to fill with each other in the region of the vortices. Similar case occurs in the molecular BECs with the vortices of the $m_{F}=0$ component locating at the center.
The atom-molecule conversion forces the atomic vortex of spin-up (spin-down) component to couple with two molecular ones of $m_{F}=+1$ ($m_{F}=-1$) components, just like that in the normal atomic-molecular BECs \cite{Woo}.
This leads to the formation of the above vortices cluster.

Since our system includes the pseudo-spin-1/2 atomic BECs and the spin-1 molecular BECs,
there will be spin textures coming from atomic BECs and molecular BECs, respectively.
Thus, this mixed system will show the interaction of the spin textures induced by different hyperfine spin states.
In pseudo-spin-1/2 BECs and spin-1 BECs, the spin texture \cite{ Mizushima, shima, Kasamatsu, matsu} is defined by
\begin{eqnarray} \label{sss}
\textbf{S}^{atom}_{\alpha}&=&\sum_{m,n=\uparrow \downarrow}\psi^{*}_{m}(\widehat{\sigma}_{\alpha})_{m,n}\psi_{n}/|\mathbf{\psi}|^{2}, (\alpha=x,y,z),\notag\\
\textbf{S}^{molecule}_{\alpha}&=&\sum_{m,n=0,\pm1}\phi^{*}_{m}(\widehat{F}_{\alpha})_{m,n}\phi_{n}/|\mathbf{\phi}|^{2},(\alpha=x,y,z),
\end{eqnarray}
where $\widehat{\sigma}_{\alpha}$ and $\widehat{F}_{\alpha}$ are the Pauli and spin-1 matrices, respectively.

The spin textures in the atomic BECs and molecular BECs are shown in Figs. 2c and 2d, respectively.
In Fig. 2c, the arrows form a circle located near the position $(0, 0)$. The arrows trend to point down in the region $y<0$ and up in the region $y>0$.
Similarly, in Fig. 2d, the arrows form two half circles located near the positions $(0, -0.7)$ and $(0, 0.7)$, respectively.
Performing a transformation
$(\textbf{S}^{'}_{x}, \textbf{S}^{'}_{y}, \textbf{S}^{'}_{z})=(\textbf{S}_{z}, \textbf{S}_{y}, -\textbf{S}_{x})$,
one can identify clearly a Skyrmion [see Fig. 2e] and two half-Skyrmions [see Fig. 2f].
In the Supplementary Information, we have proved that the transformation does not affect the value of the topological charge $Q$ [$Q=\frac{1}{4\pi}\int\int q(x,y)dxdy$, where $q(x,y)=\textbf{s}\cdot(\frac{\partial \textbf{s}}{\partial x}\times\frac{\partial \textbf{s}}{\partial y})$, $\textbf{s}=\textbf{S}/|\textbf{S}|$]. Additionally, we provide approximate solutions for the carbon-dioxide-like Skyrmion [see the Supplementary Information].
Figure 2g is a combined scheme of a carbon-dioxide-like Skyrmion, where a Skyrmion of atomic BECs couples with two half-Skyrmions of molecular BECs. Figure 2h shows the differences of spin vectors between the atomic BECs and the molecular BECs ($y=0$).
It indicates that the spin vectors of atomic BECs and molecular BECs would trend to be parallel to each other except at the region of the carbon-dioxide-like Skyrmion.
Figures 2i and 2j show the topological charge density $q(x,y)$.
The positions of vortices are pointed out to highlight the relationship between Skyrmion and vortices.
We find that the topological charge approaches $-1$ and $-\frac{1}{2}$ for the two types of spin texture, respectively.
The two zero points in the Fig. 2j further prove that we obtain two half-Skyrmions.
Figure 2k shows the sum of topological charge density [$q^{sum}(x,y)=q^{atom}(x,y)+q^{molecule}(x,y)$], which further clearly indicates the carbon-dioxide-like structure.

\subsection{The effect of spin-orbit coupling on carbon-dioxide-like Skyrmions in atomic-molecular BECs.}

We now explore an effect of the SOC on the rotating atomic-molecular BECs.
Figures 3a, 3c, 3e display densities of the atomic-molecular BECs obtained for the equilibrium state at $\Omega_{rotation}=0.5\omega$.
Figures 3b, 3d, 3f demonstrate the corresponding phases.
Without the SOC ($\lambda=0$), two atomic and three molecular components of the BECs do not display the phase separation [see Fig. 3a].
Yet there are a lot of vortices in each component, and they can be identified by the phase image [see Fig. 3b].
However, adding the SOC ($\lambda=1$), the two components of atomic BECs split into two parts [see Fig. 3c].
The component described by $\psi_{\uparrow}$ is located below the $x$ axis, whereas $\psi_{\downarrow}$ is localised mainly above it.
Similarly, the phase separation also occurs in the molecular BECs.
Because the spin-up (spin-down) atoms can form the $\phi_{+1}$ ($\phi_{-1}$) component, the corresponding molecular BECs of $\phi_{+1}$ ($\phi_{-1}$) also tends to locate below (upper) $y=0$ axis.
The $\phi_{0}$ component is spread around the $x$ direction and squeezed in the $y$ direction, and the area is filled with the vortices. When the strength of the SOC is up to $\lambda=2$, the phase separation becomes more obvious [see Fig. 3e]. In fact, the phase separation does not take place at all if the rotation is absent.
Therefore, the combination of SOC and rotation induces the phase separation in the atomic-molecular BECs.

To detect the underlying relationship between the topological excitations, Figures 3g-3i show the position of vortices, where the results correspond
to Figs. 3a, 3c, and 3e, respectively. Note that we do not indicate the vortices where the densities of BECs are very low. In that case, the contribution of energy and momentum to the system almost can be neglected.
Without the SOC [see Fig. 3g] five types of vortices form certain clusters (a black rectangle shows an example).
The repulsion among vortices of each component forces them to spread over the BECs, but they do not form a regular lattice.
At the same time, the repulsion among vortex clusters also leads to the deformation of the clusters shown in Fig. 3g.
In Fig. 3h ($\lambda=1$), a black rectangle shows some regular structure of vortices occurring around the region along the $x$-axis.
In fact, they are essentially vortex clusters which degenerate into a line. The number of vortices in each column is the same as that in Fig. 2a.
The SOC leads the vortices to occur along the $x$ axis, but the repulsion among vortices squeezes them to distribute in a line. Thus, several lines of vortices form a 2D lattice.
In a periphery of the lattice there are the remaining vortex clusters. Outside the BECs, the vortices fail to form the full vortex clusters, because the phase separation causes some components to disappear locally.
When the strength of the SOC is increased up to $\lambda=2$, the structure of the vortices lattice (a black rectangle region in Fig. 3i) is different from that in Fig. 3h. Here, each vortex of $\phi_{0}$ component tends to overlap with an atomic vortex. And the vortex number of $\phi_{+1}$ ($\phi_{-1}$) component decreases to 1 in each column.
The blue rectangle and the red rectangle in Fig. 3i indicate that vortices form some carbon dioxide structures further away form the center of the vortex lattice (black rectangle).
Above the black rectangle, vortices in components of $\psi_{\downarrow}$ and $\phi_{-1}$ form the carbon dioxide structure, whereas the atomic vortex are situated in the middle part, and two molecular vortices are located at the two sides symmetrically.
Similarly, the carbon dioxide structure of vortices in components of $\psi_{\uparrow}$ and $\phi_{+1}$ occurs below the black rectangle.
These results indicate that the strong SOC deepens the phase separation.
At the same time, it forces the vortices cluster to transform into the lattice around the $x$ axis and even into the carbon dioxide structures in the remaining regions.

We now discuss the effect of the SOC on the spin texture on the above results.
Figure 4 shows the spin texture after the transformation $(\textbf{S}^{'}_{x}, \textbf{S}^{'}_{y}, \textbf{S}^{'}_{z})=(\textbf{S}_{z}, \textbf{S}_{y}, -\textbf{S}_{x})$.
In the Supplementary Information, it has been proved that the transformation only changes the formation of the spin texture but does not change its character.
Figures 4a and 4d show the spin texture in atomic BECs and molecular BECs without SOC ($\lambda=0$), respectively.
The topological charge density in Fig. 4g (Fig. 4j) shows that they are Skyrmion (half-Skyrmion) with $Q=-1$ ($Q=-\frac{1}{2}$).
The position of vortices proves that the Skyrmion (half-Skyrmion) originates from the vortices dipole of atomic BECs (the three vortices structure of molecular BECs).
The sum of topological charge density in Fig. 4m indicates that the Skyrmion of atomic BECs and the half-Skyrmion of molecular BECs
form the carbon dioxide structure, which is marked out by blue ellipses.

The SOC can redistribute the carbon-dioxide-like Skyrmions in the atomic-molecular BECs.
For $\lambda=1$, we also obtain some Skyrmions and half-Skyrmions in the atomic BECs and molecular BECs, respectively. Due to the phase separation caused by the SOC and rotation, some Skyrmions are forced to distribute along the $x$-axis in Fig. 4b. Similarly, some half-Skyrmions distribute parallel along the $x$-axis in Fig. 4e.
The topological charge density in Figs. 4h, 4k and 4n shows the carbon-dioxide-like Skyrmions distribute in a line along the $x$-axis.

The strong SOC can even redesign the spin excitations of the textures.
When the strength of the SOC $\lambda$ reaches 2, the Skyrmion can occur in the atomic BECs [see Figs. 4c and 4i]. The Skyrmion along the $x$-axis is induced by the vortices dipole but away from the $x$-axis they are induced by a single vortex.
In the molecular BECs, the half-Skyrmion is induced by a single vortex of $\phi_{-1}$ component in the region of $y>1.5$ [see Figs. 4f and 4l].
Along the $x$ axis (between the black lines), the spin texture transforms into a structure with a fractional topological charge.
This structure is not constructed by the normal half-Skyrmion because the singular points in the topological charge densities disappear.
The sum of the topological charge density in Fig. 4o further proves that the carbon-dioxide-like Skyrmion indeed occur in the region where $y>1.5$ or $y<-1.5$.

\subsection{The effect of atom-molecule conversion on the carbon-dioxide-like Skyrmion in the spin-orbit coupling atomic-molecular BECs.}

The effect of atom-molecule conversion on the spin textures in atomic-molecular BECs is shown in Figs. 5a-5f.
We find that it is hard to mark out the carbon-dioxide-like Skyrmion after the atom-molecule conversion being terminated (i.e., $\chi=0.0$, see Figs. 5a and 5d).
The topological charge densities in Figs. 5g and 5j prove that lots of Skyrmions and half-Skyrmions are induced. However, without the atom-molecule conversion, the two kinds of spin textures do not couple with each other. The sum of the two topological charge densities in Fig. 5m further shows that it is hard to separate these structures into independent unities, which can be marked out by the blue ellipses.
When the atom-molecule conversion is $\chi=0.01$, we can divide generally the topological charge densities into several independent unities by the blue ellipses (see Figs. 5h, 5k and 5n). This kind of dividing also is shown in Figs. 5b and 5e. Clearly, these structures are carbon-dioxide-like Skyrmions. In fact, we find some ellipses would overlap partly. When the atom-molecule conversion is up to $\chi=0.02$, there is no any overlapping among the dividing (see Figs. 5c, 5f, 5i, 5l and 5o).
Thus, the atom-molecule conversion, which is the key point of the superchemistry effect, induces the coupling between Skyrmion in atomic BECs and half-Skyrmion in molecular BECs. Therefore, the carbon-dioxide-like Skyrmion is a special phenomenon in atomic-molecular BECs because the atom-molecule conversion is a needed factor to induce the coupling.

In the Supplementary Information, we indicate the corresponding densities distribution of the atomic-molecular BECs [see Figs. S1a-S1c], and the positions of vortices [see Figs. S1d-S1l].

\section*{Discussion}

We have explored the formation of the carbon-dioxide-like Skyrmions, where one Skyrmion of atomic BECs couples with two half-Skyrmions coming from molecular BECs to form the carbon-dioxide structure in the SOC atomic-molecular BECs.
Our results also show systematically the effect of SOC and atom-molecule conversion on the creation of this spin textures excitation.
This kind of Skyrmions originate from some vortices structures such as the vortex cluster and the carbon-dioxide atomic-molecular vortices.
SOC can make the vortex clusters favor to distribute along the $x$ axis, and the repulsion among vortex clusters squeezes them into lines.
Thus, the vortices lattice tends to occur around the $x$ axis.
Meanwhile, the carbon-dioxide-like Skyrmions are forced to distribute in parallel along the $x$ axis due to the effect of SOC on the vortices.
Very strong SOC changes the structure of vortices lattice and destroys the half-Skyrmion around the $x$ axis. But it also makes the vortices cluster degenerate into carbon dioxide vortices, which can induce the carbon-dioxide-like Skyrmion.
Skyrmions are often viewed as the ideal information carriers of great prospect in future application via the creation and annihilation \cite{Romming, Nagaosa}.
The carbon-dioxide-like Skyrmion is the result of combining the separated topological excitations to form a structural cell. This property implies that if the carbon-dioxide-like Skyrmions can serve as information carriers \cite{Romming, Nagaosa}, including the creation and annihilation, their distribution in lines and in parallels, and the combination and separation will demonstrate the potential for
information storage. Thus, the carbon-dioxide-like Skyrmion allows more complex control than normal Skyrmion discussed previously for information carriers.

In the Supplementary Information, we further prove that the Raman coupling enhances the creation of carbon-dioxide-like Skyrmion (see Supplementary Figs. S2, S3 and S4). The case in Fig. 3(a) of single Raman coupling without SOC would be realized by lasers with the same direction. It is impossible to only hold the Raman coupling but neglect the SOC with the counter-propagating Raman lasers. If the counter-propagating Raman lasers are turned off, i.e., both Raman coupling and SOC are terminated, the carbon-dioxide-like Skyrmions do not occur in the rotating atomic-molecular BECs although the superchemistry effect is considered (see Supplementary Figs. S5).
Furthermore, there are no nontrivial spin texture excitations if the rotation frequency decreases to be zero.
Due to the dependence of the spin-dependent photoassociation and Raman lasers, the carbon-dioxide-like Skyrmion is a special phenomenon existing in the SOC atomic-molecular BECs with hyperfine spin states.

In real experiment, the carbon-dioxide-like Skyrmion can be identified by a direct observation of the densities via time-of-flight absorption imaging technique \cite{Leanhardt, Leslie, Choi}, which can measure the characteristics of the carbon-dioxide-like Skyrmion by means of corresponding quantum vortices.
The experiments should begin with nearly pure $^{87}$Rb BEC of approximately $3.0\times10^{5}$ or more
atoms in the $|f=1,m_{f}=-1\rangle$ state \cite{Lind, Shuai-Chen2, Williams, Hamner, ChenYong, Zhangchuanwei}.
An initially off-resonant radio-frequency magnetic field can be used to prepare equal mixtures of $|f=1,m_{f}=-1\rangle$ and $|f=1,m_{f}=0\rangle$.
Then, the spin-dependent photoassociation of atom \cite{Hamley, Kobayashi} creates the molecules of $|F=2,m_{F}=0,-1,-2\rangle$, which are used to produce the excited molecules of $|F=1,m_{F}=0,\pm1\rangle$ via the transition $\Delta F=-1$ and $\Delta m_{F}=+1$, and finally the excited molecules degenerate to be the molecular BECs of $|F=1,m_{F}=0,\pm1\rangle$ [see Figs. 1d-1f]. At the same time, the counter-propagating Raman lasers $L3$ and $L4$ ($L5$, $L6$ and $L7$) shown in Fig. 1 are used to induce SOC for atomic (molecular) BECs \cite{Lind, Shuai-Chen2, Williams, Hamner, ChenYong, Zhangchuanwei}.
Combining rotation effect and the further cooling process, one would obtain the SOC atomic-molecular BECs at the equilibrium state.
In this study, our results have shown that the carbon-dioxide-like Skyrmion is related either to vortices cluster or to carbon-dioxide vortices. The vortex cluster consists of two atomic vortices and six molecular vortices [see the black rectangle in Fig. 3g]. The structure of the carbon-dioxide vortices is simple [see the red and blue rectangles in Fig. 3i].
Therefore, one can apply a ballistic expansion (approximate 20ms) of the BECs in a gradient magnetic field to separate the
different spin states \cite{Leanhardt, Leslie, Choi}. Vortices will be imprinted by ramping the magnetic field $B_{z}\longrightarrow0$.
After the structures of vortices are measured, the carbon-dioxide-like Skyrmion can be identified.

In summary, we have constructed an experimental proposal for creating SOC in atomic-molecular BECs by combining the spin dependent photoassociation and Raman coupling.
This type of SOC will create the novel topological excitation -- carbon-dioxide-like Skyrmion and also control the structures of this Skyrmions, which can be measured by the time-of-flight absorption imaging technique.
This work opens a new window to realize of SOC in multi-component ultracold gases and manipulate complex nontrivial topological excitations in future experiments.

\section*{Methods}

\subsection{The equations of the spin-orbit coupling atomic-molecular BECs.}

The SOC of atomic BECs can be generated by two additional
Raman beams ($\omega_{L3}$ and $\omega_{L4}$) which couple the atomic internal states $|\uparrow\rangle$
and $|\downarrow\rangle$. The single-atom Hamiltonian induced by
the Raman coupling is $\Omega_{A}|\uparrow\rangle\langle\downarrow|e^{-2ik_{x}x}+h.c.$,
with $\Omega_{A}$ the atomic Raman coupling intensity.
Furthermore, Raman beams ($\omega_{L5}$, $\omega_{L6}$ and $\omega_{L7}$) couple different molecular internal states
$|m_{F}=0,\pm1\rangle$, and lead to Hamiltonain
$\left(|m_{F}=
0\rangle
\langle m_{F}=-1|+
|m_{F}=1\rangle\langle m_{F}=0|\right)\Omega_{M}e^{ik_{x}X}+h.c.$,
with $X$ the position of the mass center of the two atoms and $\Omega_{M}=\sqrt{2}\Omega_{A}$.

To clearly illustrate effects of the SOC, we introduce
a unitary transformation
$U_{A}=e^{i\sigma_{z}k_{x}x}$ and $U_{M}=\sum_{m=0,\pm1}e^{2imk_{x}X}|m_{F}=m\rangle\langle m_{F}=m|$
for the atomic states and molecular states, respectively, with
$\sigma_{z}=|\uparrow\rangle\langle\uparrow|-|\downarrow\rangle\langle\downarrow|$.
Therefore, in the spin rotated frame induced by a combined transformation
$U=U_{A}P^A+U_{M}P^M$, the Hamiltonian is $H=UH'U^{\dagger}$, where $H'$
is the Hamiltonian in the original (un-rotated) frame, $P^A$ and $P^M$ being the projection operators onto
the atomic and molecular states.
The straightforward
calculation shows that,
for the condensate in the $d$-dimensional
space ($d=1,2,3$), in the second quantization we have
\begin{equation}
H=H_{AF}+H_{MF}+H_{SOC}+H_{Raman}+H_{AM}+H_{I}.\label{h}
\end{equation}
Here $H_{AF}=\sum_{a=\uparrow,\downarrow}\int\hat{\psi}_{a}^{\dagger}(\vec{r})[-\hbar^{2}\nabla_{d}^{2}/(2M_{A})+M_{A}\omega^{2}r^{2}/2]\hat{\psi}_{a}(\vec{r})d\vec{r}$
and $H_{MF}=
\sum_{m=0,\pm1}\int\hat{\phi}_{m}^{\dagger}\\
(\vec{r})
[-\hbar^{2}\nabla_{d}^{2}/(4M_{A})+M_{A}\omega^{2}r^{2}+\varepsilon+M_{A}\lambda^{2}m^{2}]\hat{\phi}_{m}(\vec{r})d\vec{r}$
are the free Hamiltonians for atoms and molecules, respectively, where
$M_{A}$ is the single-atom mass, $\omega$ is the trapping frequency,
$\varepsilon$ is the difference between the binding energy to the
two-atom molecule and the frequency difference of the two laser beams
which induce atom-molecule coupling, the factor $\lambda$ is defined
as $\lambda=k_{x}/M_{A}$, the operator $\hat{\psi^{\dagger}}_{a}(\vec{r})$
($a=\uparrow,\downarrow$) is the creation operator for a single atom
at position $\vec{r}$ with internal state $|a\rangle$, and $\hat{\phi}_{m}^{\dagger}(\vec{r})$
($m=0,\pm1$) is the creation operator for a single molecular at position
$\vec{r}$ with internal state $|m_{F}=m\rangle$. In Eq. (\ref{h})
the SOC is described by
\begin{eqnarray}
H_{SOC}=\lambda\int\hat{\psi}_{\uparrow}^{\dagger}(\vec{r})(-i\hbar\partial_{x})\hat{\psi}_{\uparrow}(\vec{r})d\vec{r}-
\lambda\int\hat{\psi}_{\downarrow}^{\dagger}(\vec{r})(-i\hbar\partial_{x})\hat{\psi}_{\downarrow}(\vec{r})d\vec{r}\notag\\
+\lambda\int\hat{\phi}_{1}^{\dagger}(\vec{r})(-i\hbar\partial_{x})\hat{\phi}_{1}(\vec{r})d\vec{r}-\lambda\int\hat{\phi}_{-1}^{\dagger}(\vec{r})(-i\hbar\partial_{x})\hat{\phi}_{-1}(\vec{r})d\vec{r}.\notag
\end{eqnarray}
The Ramman coupling between internal states is described by
\begin{eqnarray}
H_{Raman}=
\Omega_{A}\int\hat{\psi}_{\uparrow}^{\dagger}(\vec{r})\hat{\psi}_{\downarrow}(\vec{r})d\vec{r}+\Omega_{M}\int\hat{\phi}_{1}^{\dagger}(\vec{r})\hat{\phi}_{0}(\vec{r})d\vec{r}+\Omega_{M}\int\hat{\phi}_{0}^{\dagger}(\vec{r})\hat{\phi}_{-1}(\vec{r})d\vec{r}+h.c..\notag
\end{eqnarray}
The laser-induced atom-molecule transition is described by
\begin{eqnarray}
H_{AM}=\chi\int d\vec{r}[\hat{\psi}_{1}^{\dagger}(\vec{r})\hat{\phi}_{\uparrow}^{2}(\vec{r})/\sqrt{2}+\hat{\psi}_{0}^{\dagger}(\vec{r})\hat{\phi}_{\uparrow}(\vec{r})\hat{\phi}_{\downarrow}(\vec{r})+\hat{\psi}_{-1}^{\dagger}(\vec{r})\hat{\phi}_{\downarrow}^{2}(\vec{r})/\sqrt{2}+h.c.].\notag
\end{eqnarray}
In our system the inter-particle interaction is described by $H_{I}$
in Eq. (\ref{h}), and we have $H_{I}=U_{AA}+U_{AM}+U_{MM}$. Here
$U_{AA}=\sum_{a=\uparrow,\downarrow}g_{aa}\int d\vec{r}\hat{\psi}_{a}^{\dagger}(\vec{r})^{2}\hat{\psi}_{a}(\vec{r})^{2}/2+g_{\uparrow\downarrow}\int d\vec{r}\hat{\psi}_{\uparrow}^{\dagger}(\vec{r})\hat{\psi}_{\downarrow}^{\dagger}(\vec{r})\hat{\psi}_{\downarrow}(\vec{r})\hat{\psi}_{\downarrow}(\vec{r})$
describes the atom-atom collision, $U_{AM}=
\sum_{a=\uparrow,\downarrow}\sum_{m=0,\pm1}g_{am}
\int d\vec{r}\hat{\psi}_{a}^{\dagger}(\vec{r})\hat{\phi}_{m}^{\dagger}(\vec{r})\hat{\phi}_{m}(\vec{r})\hat{\psi}_{a}(\vec{r})$
describes the atom-molecule collision, and $U_{MM}$ describes the
molecule-molecule collision. In usual case the collision between molecules
with different internal states are quite complicated. Here for simplicity
we assume in our system the scattering of three-component molecules
has the same property as the scattering of $F=1$ atoms \cite{HoTLspinor, Ohmispinor}, i.e., the
scattering amplitude is totally determined by the total spin of the
two molecules. In that case we have
$U_{MM}=\sum_{m,m'=0,\pm1}\int d\vec{r} [\frac{g_{n}}{2}\hat{\phi}_{m}^{\dagger}(\vec{r})\hat{\phi}_{m'}^{\dagger}(\vec{r})\\
\hat{\phi}_{m'}(\vec{r})\hat{\phi}_{m}(\vec{r})]
+
\sum_{\alpha=x,y,z}\sum_{m,n,k,l=0\pm1}
\int d\vec{r}[\frac{g_{s}}{2}\hat{\phi}_{k}^{\dagger}(\vec{r})\hat{\phi}_{n}^{\dagger}(\vec{r})(F_{\alpha})_{nm}
(F_{\alpha})_{kl}\hat{\phi}_{m}(\vec{r})\hat{\phi}_{l}(\vec{r})]$.

In this way, we obtain the coupled equations of the atomic-molecular BECs with SOC. It can be written as
\begin{eqnarray}
i\hbar\frac{\partial\psi_{\uparrow}}{\partial t}&=&[-\frac{\hbar^{2}\nabla_{d}^{2}}{2M_{A}}+\frac{M_{A}\omega^{2}r^{2}}{2}]\psi_{\uparrow}+\lambda p_{x}\psi_{\uparrow}+\hbar\Omega_{A}\psi_{\downarrow}\notag\label{GP2}\\
&+&(\sum_{a=\uparrow,\downarrow}g_{\uparrow,a}|\psi_{a}|^{2}+\sum_{m=0,\pm1}g_{\uparrow,m}|\phi_{m}|^{2})\psi_{\uparrow}+(\sqrt{2}\chi\psi_{\uparrow}^{*}\phi_{+1}+\chi\psi_{\downarrow}^{*}\phi_{0}),
\end{eqnarray}
\begin{eqnarray}
i\hbar\frac{\partial\psi_{\downarrow}}{\partial t}&=&[-\frac{\hbar^{2}\nabla_{d}^{2}}{2M_{A}}+\frac{M_{A}\omega^{2}r^{2}}{2}]\psi_{\downarrow}-\lambda p_{x}\psi_{\downarrow}+\hbar\Omega_{A}\psi_{\uparrow}\notag\\
&+&(\sum_{a=\uparrow,\downarrow}g_{\downarrow,a}|\psi_{a}|^{2}+\sum_{m=0,\pm1}g_{\downarrow,m}|\phi_{m}|^{2})\psi_{\downarrow}+(\sqrt{2}\chi\psi_{\downarrow}^{*}\phi_{-1}+\chi\psi_{\uparrow}^{*}\phi_{0}),
\end{eqnarray}
\begin{eqnarray}
i\hbar\frac{\partial\phi_{+1}}{\partial t}=[-\frac{\hbar^{2}\nabla_{d}^{2}}{2M_{M}}+\frac{M_{M}\omega^{2}r^{2}}{2}]\phi_{+1}+\frac{\chi}{\sqrt{2}}\psi_{\uparrow}^{2}+(\varepsilon+M_{A}\lambda^{2})\phi_{+1}+\lambda p_{x}\phi_{+1}+\sqrt{2}\hbar\Omega_{A}\phi_{0}\notag\\
+(\sum_{a=\uparrow,\downarrow}g_{a,+1}|\psi_{a}|^{2}+g_{n}\sum_{m=0,\pm1}|\phi_{m}|^{2})\phi_{+1}+g_{s}(|\phi_{+1}|^{2}+|\phi_{0}|^{2}-|\phi_{-1}|^{2})\phi_{+1}+g_{s}\phi_{-1}^{+}\phi_{0}\phi_{0},
\end{eqnarray}
\begin{eqnarray}
i\hbar\frac{\partial\phi_{0}}{\partial t}=[-\frac{\hbar^{2}\nabla_{d}^{2}}{2M_{M}}+\frac{M_{M}\omega^{2}r^{2}}{2}]\phi_{0}+\chi\psi_{\uparrow}\psi_{\downarrow}+\varepsilon\phi_{0}+\sqrt{2}\hbar\Omega_{A}(\phi_{+1}+\phi_{-1})\notag\\
+(\sum_{a=\uparrow,\downarrow}g_{a,0}|\psi_{a}|^{2}+g_{n}\sum_{m=0,\pm1}|\phi_{m}|^{2})\phi_{0}+g_{s}(|\phi_{+1}|^{2}+|\phi_{-1}|^{2})\phi_{0}+2g_{s}\phi_{0}^{+}\phi_{+1}\phi_{-1},
\end{eqnarray}
\begin{eqnarray}
i\hbar\frac{\partial\phi_{-1}}{\partial t}=[-\frac{\hbar^{2}\nabla_{d}^{2}}{2M_{M}}+\frac{M_{M}\omega^{2}r^{2}}{2}]\phi_{-1}+\frac{\chi}{\sqrt{2}}\psi_{\downarrow}^{2}+(\varepsilon+M_{A}\lambda^{2})\phi_{-1}-\lambda p_{x}\phi_{-1}+\sqrt{2}\hbar\Omega_{A}\phi_{0}\notag\\
+(\sum_{a=\uparrow,\downarrow}g_{a,-1}|\psi_{a}|^{2}+g_{n}\sum_{m=0,\pm1}|\phi_{m}|^{2})\phi_{-1}+g_{s}(|\phi_{-1}|^{2}+|\phi_{0}|^{2}-|\phi_{+1}|^{2})\phi_{-1}+g_{s}\phi_{+1}^{+}\phi_{0}\phi_{0},
\end{eqnarray}

\subsection{The damped projected Gross-Pitaevskii equation.}
The damped projected Gross-Pitaevskii equation (PGPE) \cite{ Rooney} is used to obtain the ground state of atomic-molecular BECs.
By neglecting the noise term according to the corresponding stochastic PGPE \cite{ Bradley},
the damped PGPE for the atomic-molecular BECs is described as
\begin{eqnarray}
d\psi_{a}=\mathcal{P}\{-\frac{i}{\hbar}\widehat{H}_{a}\psi_{a}dt+\frac{\gamma_{a}}{k_{B}T}(\mu_{a}-\widehat{H}_{a})\psi_{a}dt\},\\
d\phi_{m}=\mathcal{P}\{-\frac{i}{\hbar}\widehat{H}_{m}\phi_{m}dt+\frac{\gamma_{m}}{k_{B}T}(\mu_{m}-\widehat{H}_{m})\phi_{m}dt\},
\end{eqnarray}
where, $\widehat{H}_{a}\psi_{a}=i\hbar\frac{\partial\psi_{a}}{\partial t}$ ($a=\uparrow,\downarrow$), $\widehat{H}_{m}\phi_{m}=i\hbar\frac{\partial\phi_{m}}{\partial t}$ ($m=0, \pm 1$), $T$ is the final temperature, $k_{B}$ is the Boltzmann constant, $\mu_{a}$ and $\mu_{m}$ are the chemical potential of atom and molecule respectively. $\gamma_{a}$ ($\gamma_{m}$) is the growth rate for the $a$th component ($m$th component). To obtain the ground state, we set the parameter $\frac{\gamma_{a}}{k_{B}T}=\frac{\gamma_{m}}{k_{B}T}=0.05$ directly. The projection operator $\mathcal{P}$ is used to restrict the dynamics of atomic-molecular BEC in the coherent region.
The initial state of each component is generated by sampling the grand canonical
ensemble for a free ideal Bose gas with the chemical potential $\mu_{m,0}=2\mu_{a,0}$. The final chemical potential of the noncondensate band are altered to the values $\mu_{m}=2\mu_{a}$.



\begin{addendum}

\item [Acknowledgement]
We would like to thank P. Zhang and H. Jing for helpful discussions.
C. F. L. is supported by the NSFC under grants Nos. 11304130, 11365010,
and Postdoctoral Science Foundation of China (Grant No. 2014M550868).
W. M. L. is supported by the NSFC under grants Nos. 11434015, 61227902, 61378017,
NKBRSFC under grants Nos. 2011CB921502, 2012CB821305, SKLQOQOD under grants No. KF201403, SPRPCAS under grants No. XDB01020300.


\item [Author Contributions]

W.M.L. conceived the idea and supervised the overall research. C.F.L. designed and
performed the numerical experiments. G. J. developed the theory. C.F.L. wrote the
paper with helps from all other co-authors.

\item [Competing Interests]
The authors declare that they have no competing financial interests.

\item [Correspondence]
Correspondence and requests for materials should be addressed to Liu, Wu-Ming.
\end{addendum}

\clearpage

\newpage
\bigskip
\textbf{Figure 1 Experimental setup for creating the spin-orbit coupled atomic-molecular BECs of $^{87}$Rb.}
(\textbf{a}) The experimental geometry. The formation of spin molecular BECs: laser field $L1$ finishes the spin dependent photoassociation \cite{Hamley}, laser field $LT$ performs the $\Delta F=-1$, $\Delta m_{F}=+1$ transiting process of the excited molecules, and laser field $L2$ induces the excited molecules to emit a photon and forms the stable spin-1 molecular BECs. The creation of SOC: the laser fields $L3$ and $L4$ ($L5$, $L6$ and $L7$) are used to create the SOC of atomic (molecular) BECs along the $x$ axis. $B$ denotes a weak magnetic field along the $z$ axis, which is used to induce the Zeeman shift.
(\textbf{b}) Level diagram of Raman coupling within the atomic BECs. Two lasers ($L3$ and $L4$) is used to couple states of $|\uparrow\rangle$ (i.e., $|1,-1\rangle$) and $|\downarrow\rangle$ (i.e., $|1,0\rangle$). This leads to the creation of the SOC in atomic BECs.
(\textbf{c}) Level diagram of Raman coupling within the $F=1$ molecular BECs. The three lasers ($L5$, $L6$ and $L7$) couple states of $\phi_{-1}$, $\phi_{0}$ and $\phi_{+1}$. So the SOC in molecular BECs is created.
(\textbf{d}) The procedure of creating $|F=1, m_{F}=-1\rangle$ state molecular BEC. A close pair of atoms from the condensate in a state $|\downarrow\rangle$ absorb a photon from laser field $L1$,
then transit to a bound excited molecular state $|F=2,m_{F}=-2\rangle$, i.e., the spin dependent photoassociation process $|1,-1\rangle\otimes|1,-1\rangle\longrightarrow|2,-2\rangle$\cite{Hamley}. Via the laser field $LT$, the excited molecule transits to $|F=1, m_{F}=-1\rangle$ state. Finally, laser field $L2$ induces the excited molecular state to emit a photon into molecular state $\phi_{-1}$ which keeps $F=1$ and $m_{F}=-1$.
(\textbf{e}) The procedure of creating $|F=1, m_{F}=0\rangle$ state molecular BEC. Under laser field $L1$, the spin dependent photoassociation process is $|1,-1\rangle\otimes|1,0\rangle\longrightarrow|2,-1\rangle$. Via laser field $LT$, the $\Delta F=-1$, $\Delta m_{F}=+1$ transiting process leads to excited molecule of $|F=1,m_{F}=0\rangle$ state. The stable molecular BEC ($F=1$, $m_{F}=0$) is obtained via an emission of photon by laser field $L2$.
(\textbf{f}) The procedure of creating $|F=1, m_{F}=+1\rangle$ state molecular BEC. The spin dependent photoassociation process is $|1,0\rangle\otimes|1,0\rangle\longrightarrow|2,0\rangle$. The stable molecular BEC with $F=1$ and $m_{F}=+1$ is obtained via the emission of photon by the laser field $L2$.

\bigskip
\textbf{Figure 2 Creating a carbon-dioxide-like Skyrmion in the SOC atomic-molecular BECs.}
(\textbf{a}) The densities of each component when the system reaches the equilibrium state. The components are pointed out by the title: $|\uparrow\rangle$, $|\downarrow\rangle$, $m_{F}=-1$, $m_{F}=0$ and $m_{F}=+1$, respectively.
The particles number $(N_{\uparrow}, N_{\downarrow}, N_{-1}, N_{0}, N_{+1})$
is $(2.56\times10^{3}, 2.98\times10^{3}, 1.72\times10^{3}, 3.22\times10^{3}, 1.59\times10^{3})$.
The strength of SOC $\lambda=0.5$, the rotation frequency $\Omega_{rotation}=0.2\omega$, where $\omega=200\times2\pi$Hz.
We set the parameters $g_{\uparrow,\uparrow}=g_{\uparrow,\downarrow}=g_{A}$ with the scattering length $a_{A}=101.8a_{B}$, $g_{\downarrow,\downarrow}=0.95g_{A}$, $g_{AM}=0.5g_{A}$, $g_{n}=2g_{A}$, $g_{s}=0.03g_{A}$, $\chi=0.08$, $\varepsilon=0$, $\Omega_{A}=0.8$ and $\mu_{m}=2\mu_{a}=3.5\mu_{m,0}=7\mu_{a,0}=28\hbar\omega$.
(\textbf{b}) The corresponding phases of each component in \textbf{a}.
(\textbf{c}, \textbf{d}) The spin texture of the atomic BECs and molecular BECs, respectively. The color of each arrow indicates the magnitude of $S_{z}$.
(\textbf{e}, \textbf{f}) The corresponding spin texture under the transformation: $(\textbf{S}^{'}_{x}, \textbf{S}^{'}_{y}, \textbf{S}^{'}_{z})=(\textbf{S}_{z}, \textbf{S}_{y}, -\textbf{S}_{x})$.
(\textbf{g}) A combined scheme of the carbon-dioxide-like Skyrmion. A Skyrmion (center) couples with two half-Skyrmions, which have to arrange at the two sides of the Skyrmion.
(\textbf{h}) The difference of spin vectors between the atomic BECs and the molecular BECs, where $y=0$.
(\textbf{i}, \textbf{j}) The topological charge density of the atomic BECs and molecular BECs, respectively. The ellipse and rectangles mark out the regions of a Skyrmion and two half-Skyrmions, respectively.
The symbol $\circ$, $\star$, $\triangleright$, $\ast$ and $\triangleleft$ are the position of vortices formed by the $\uparrow$, $\downarrow$, $m_{F}=-1$, $m_{F}=0$ and $m_{F}=+1$ components,
respectively.
(\textbf{k}) The sum of the topological charge densities. It denotes the creation of carbon-dioxide-like Skyrmion that the ellipse overlaps with the two rectangles.
The unit of length, strength of $\chi$, $\Omega_{A}$ and SOC are $0.76\mu m$, $\hbar\omega$, $\omega$, and $0.96\times10^{-3}m/s$, respectively.

\bigskip
\textbf{Figure 3 The effect of SOC on the vortices in rotating atomic-molecular BECs.}
The rotation frequency is $\Omega_{rotation}=0.5\omega$, where $\omega=200\times2\pi$Hz.
We set the parameters $g_{\uparrow,\uparrow}=g_{\uparrow,\downarrow}=g_{A}$ with the scattering length $a_{A}=101.8a_{B}$, $g_{\downarrow,\downarrow}=0.95g_{A}$, $g_{AM}=0.5g_{A}$, $g_{n}=2g_{A}$, $g_{s}=0.03g_{A}$, $\chi=0.08$, $\varepsilon=0$, $\Omega_{A}=0.8$ and $\mu_{m}=2\mu_{a}=3.5\mu_{m,0}=7\mu_{a,0}=42\hbar\omega$.
(\textbf{a}) Densities of each components under the equilibrium state, the strength of SOC $\lambda=0$. Note that the components are pointed out by the title: $|\uparrow\rangle$, $|\downarrow\rangle$, $m_{F}=-1$, $m_{F}=0$ and $m_{F}=+1$, respectively.
(\textbf{b}) The corresponding phases of each components in (\textbf{a}).
(\textbf{c}, \textbf{d}) The corresponding densities and phases with $\lambda=1$. (\textbf{e}, \textbf{f}) The corresponding densities and phases with $\lambda=2$.
(\textbf{g}-\textbf{i}) The position of vortices in (\textbf{a}), (\textbf{c}) and (\textbf{e}), respectively. In (\textbf{g}), the rectangle points out a vortex cluster. In (\textbf{h}), the rectangle points out the region where the vortices form lattice. In (\textbf{i}), the red (blue) rectangle marks out a carbon dioxide vortex, and the black rectangle points out the vortices lattice.
The particles number $(N_{\uparrow}, N_{\downarrow}, N_{-1}, N_{0}, N_{+1})$
for the three cases are $(5.82\times10^{3}, 8.01\times10^{3}, 4.96\times10^{3}, 8.23\times10^{3}, 4.93\times10^{3})$,
$(7.86\times10^{3}, 8.80\times10^{3}, 7.95\times10^{3}, 3.93\times10^{3}, 7.85\times10^{3})$
and $(11.42\times10^{3}, 12.29\times10^{3}, 10.03\times10^{3}, 2.00\times10^{3}, 10.04\times10^{3})$, respectively.
The unit of length, strength of $\chi$, $\Omega_{A}$ and SOC are $0.76\mu m$, $\hbar\omega$, $\omega$, and $0.96\times10^{-3}m/s$, respectively.

\bigskip
\textbf{Figure 4 The effect of SOC on the spin texture in the atomic-molecular BECs.} (\textbf{a}-\textbf{c}) The spin textures of atomic BECs in Figs. 3\textbf{a}, 3\textbf{c} and 3\textbf{e}, respectively. (\textbf{d}-\textbf{f}) The spin textures of molecular BECs in Figs. 3\textbf{a}, 3\textbf{c} and 3\textbf{e}, respectively.
Note that the spin textures are under the transformation: $(\textbf{S}^{'}_{x}, \textbf{S}^{'}_{y}, \textbf{S}^{'}_{z})=(\textbf{S}_{z}, \textbf{S}_{y}, -\textbf{S}_{x})$.
The color of each arrow indicates the magnitude of $S^{'}_{z}$. The blue ellipses point out the regions of the carbon-dioxide-like Skyrmion that Skyrmion of atomic BECs couples two half-Skyrmion of molecular BECs.
In addition, we mark the position of vortices in order to illuminate
the relationship between spin texture and position of vortices clearly. The meanings of the marks are the same as that in Figs. 3\textbf{g}-3\textbf{i}.
(\textbf{g}-\textbf{i}) The corresponding topological charge density in Figs. 4\textbf{a}-4\textbf{c}, respectively. (\textbf{j}-\textbf{l}) The corresponding topological charge density in Figs. 4\textbf{d}-4\textbf{f}, respectively. In addition, we mark the position of vortices in $x<0$ region in order to illuminate the relationship between spin texture and position of vortices clearly. The meanings of the marks are the same as that in Figs. 3\textbf{g}-3\textbf{i}. (\textbf{m}-\textbf{o}) The total topological charge density of (\textbf{g}, \textbf{j}), (\textbf{h}, \textbf{k}) and (\textbf{i}, \textbf{l}), respectively.
We use blue ellipses with the same size to point out the regions of the carbon-dioxide-like Skyrmion in (\textbf{a}, \textbf{d}, \textbf{g}, \textbf{j}, \textbf{m}), (\textbf{b}, \textbf{e}, \textbf{h}, \textbf{k}, \textbf{n}) or (\textbf{c}, \textbf{f}, \textbf{i}, \textbf{l}, \textbf{o}).
The unit of length, strength of $\chi$, $\Omega_{A}$ and SOC are $0.76\mu m$, $\hbar\omega$, $\omega$, and $0.96\times10^{-3}m/s$, respectively.

\bigskip
\textbf{Figure 5 The effect of atom-molecule conversion on the spin textures in the SOC atomic-molecular BECs.}
The rotation frequency is $\Omega_{rotation}=0.5\omega$, where $\omega=200\times2\pi$Hz.
We set the parameters $\lambda=1$, $\Omega_{A}=0.8$, $g_{\uparrow,\uparrow}=g_{\uparrow,\downarrow}=g_{A}$ with the scattering length $a_{A}=101.8a_{B}$,
$g_{\downarrow,\downarrow}=0.95g_{A}$, $g_{n}=2g_{A}$, $g_{s}=0.03g_{A}$, $g_{AM}=0.5g_{A}$, $\varepsilon=0$ and $\mu_{m}=2\mu_{a}=3.5\mu_{m,0}=7\mu_{a,0}=42\hbar\omega$.
(\textbf{a}, \textbf{d}) The spin textures of atomic and molecular BECs with $\chi=0.0$.
(\textbf{b}, \textbf{e}) The spin textures of atomic and molecular BECs with $\chi=0.01$.
(\textbf{c}, \textbf{f}) The spin textures of atomic and molecular BECs with $\chi=0.02$.
The spin textures are obtained under the transformation $(\textbf{S}^{'}_{x}, \textbf{S}^{'}_{y}, \textbf{S}^{'}_{z})=(\textbf{S}_{z}, \textbf{S}_{y}, -\textbf{S}_{x})$.
The color of each arrow indicates the magnitude of $S^{'}_{z}$. In (\textbf{b}, \textbf{e}) or (\textbf{c}, \textbf{f}), we use blue ellipses with the same size to point out the regions of the carbon-dioxide-like Skyrmion. [In (\textbf{a}, \textbf{d}), we can not use the ellipses to mark out the region where Skyrmion of atomic BECs just couples two half-Skyrmions].
In addition, we mark the position of vortices in order to illuminate
the relationship between spin texture and position of vortices clearly. The meanings of the marks are the same as that in Figs. 3\textbf{g}-3\textbf{i}. The corresponding densities of the atomic-molecular BECs are shown in Figs. S1\textbf{a}-S1\textbf{c} [see the Supplementary Information].
(\textbf{g}-\textbf{l}) The corresponding topological charge densities of the spin texture in Figs. 5\textbf{a}-5\textbf{f}, respectively.
(\textbf{m}) The total topological charge densities of (\textbf{g}) and (\textbf{j}). (\textbf{n}) The total topological charge density of \textbf{h} and \textbf{k}. (\textbf{o}) The total topological charge density of \textbf{i} and \textbf{l}.
In (\textbf{b}, \textbf{e}, \textbf{h}, \textbf{k}, \textbf{n}) or (\textbf{c}, \textbf{f}, \textbf{i}, \textbf{l}, \textbf{o}), we use blue ellipses with the same size to point out the regions of the carbon-dioxide-like Skyrmion. [In (\textbf{a}, \textbf{d}, \textbf{g}, \textbf{j}, \textbf{m}), we can not use the ellipses to mark out the region where Skyrmion of atomic BECs just couples two half-Skyrmions].
In addition, we only mark the position of vortices in the region of $x<0$.
The unit of length, strength of $\chi$, $\Omega_{A}$ and SOC are $0.76\mu m$, $\hbar\omega$, $\omega$, and $0.96\times10^{-3}m/s$, respectively.

\newpage

\begin{figure}
\begin{center}
\epsfig{file=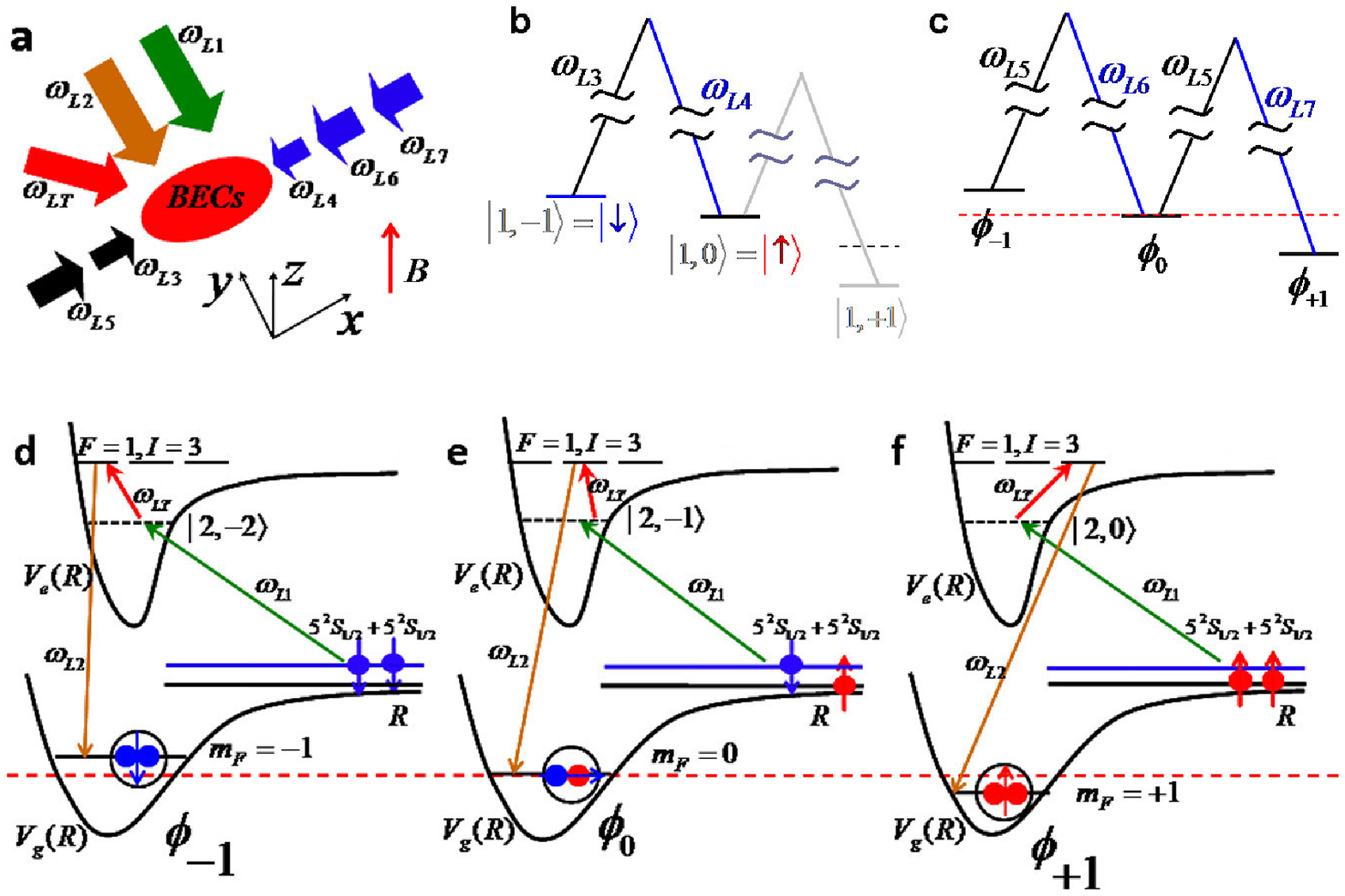,width=13cm, trim=0.0in 0.0in 0.0in 0.0in}\label{fig:scheme}
\end{center}
\end{figure}

\begin{figure}
\begin{center}
\epsfig{file=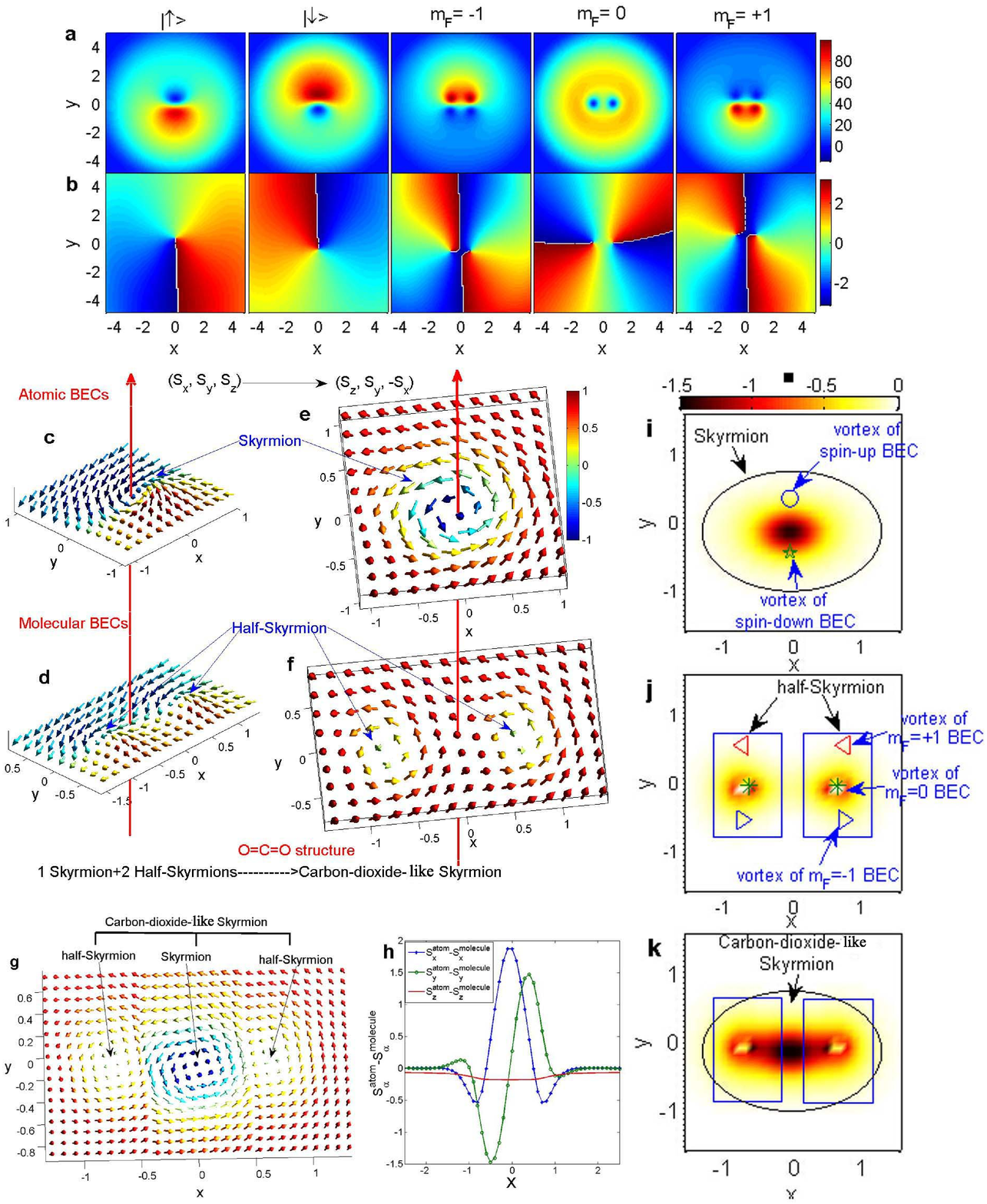,width=14cm, trim=0.0in 0.0in 0.0in 0.0in}
\end{center}
\label{figC02}
\end{figure}

\begin{figure}
\begin{center}
\epsfig{file=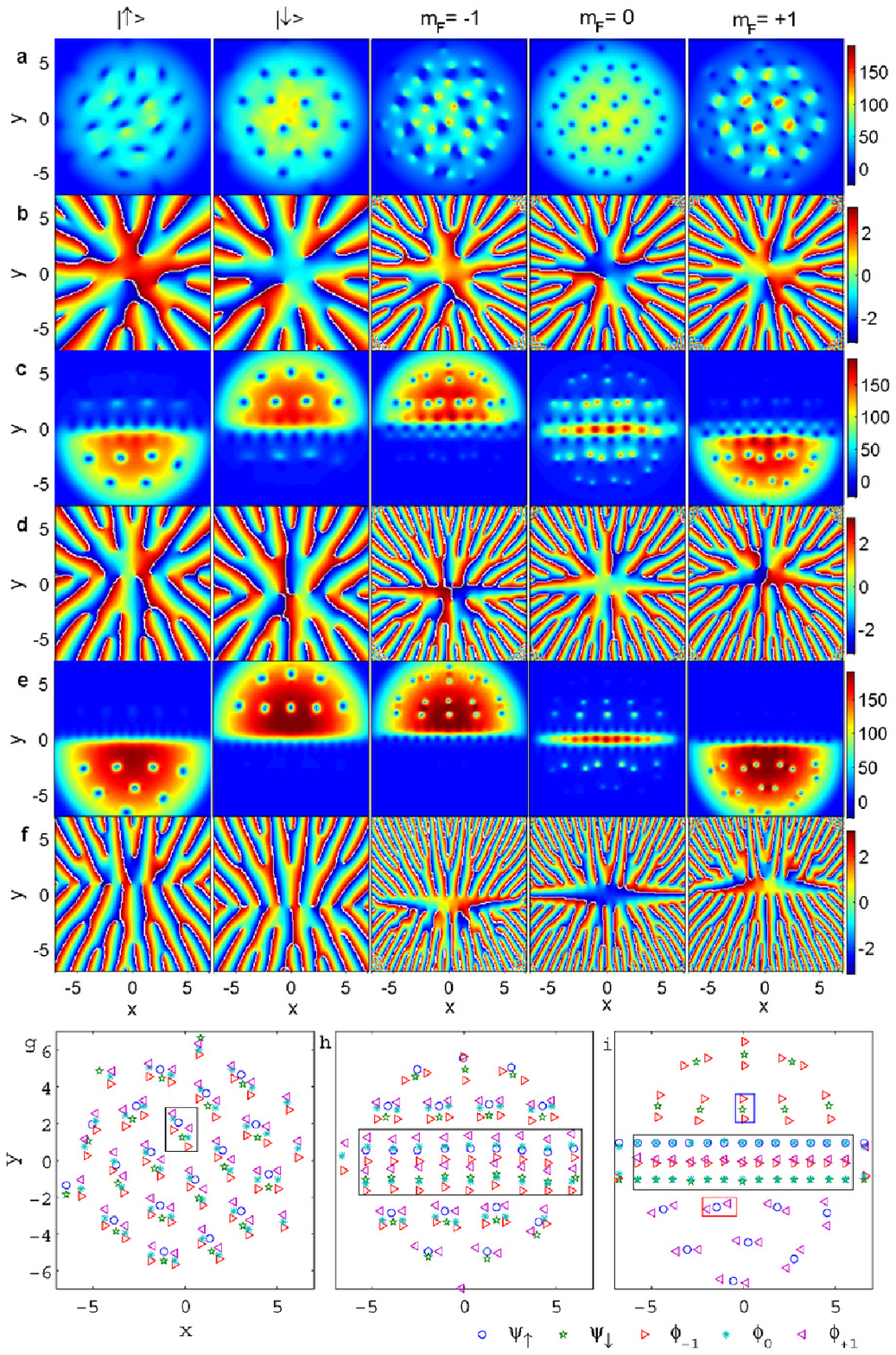,width=14.0cm}
\end{center}
\label{fig3}
\end{figure}

\begin{figure}
\begin{center}
\epsfig{file=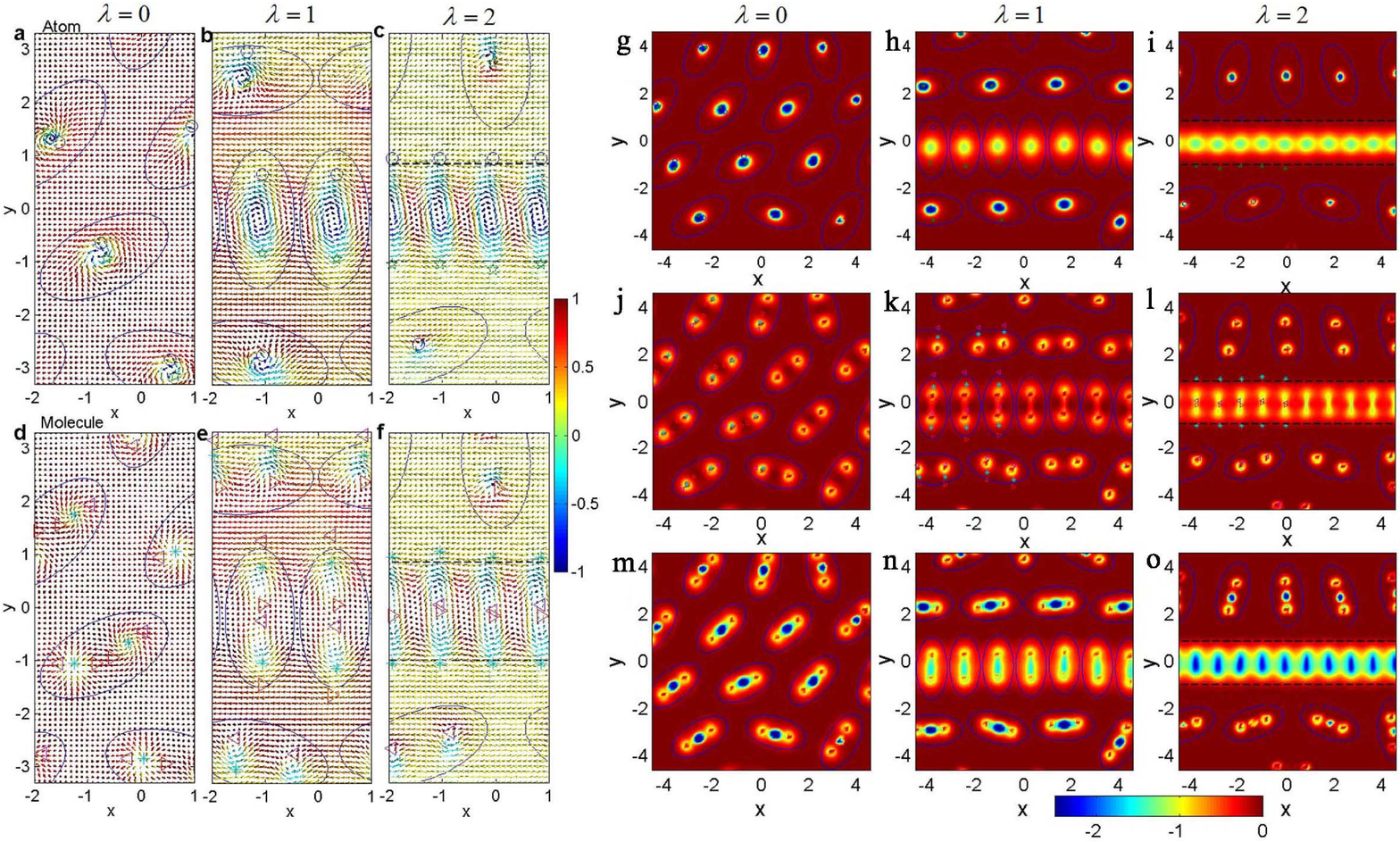,width=16.0cm}
\end{center}
\label{fig3}
\end{figure}

\newpage
\begin{figure}
\begin{center}
\epsfig{file=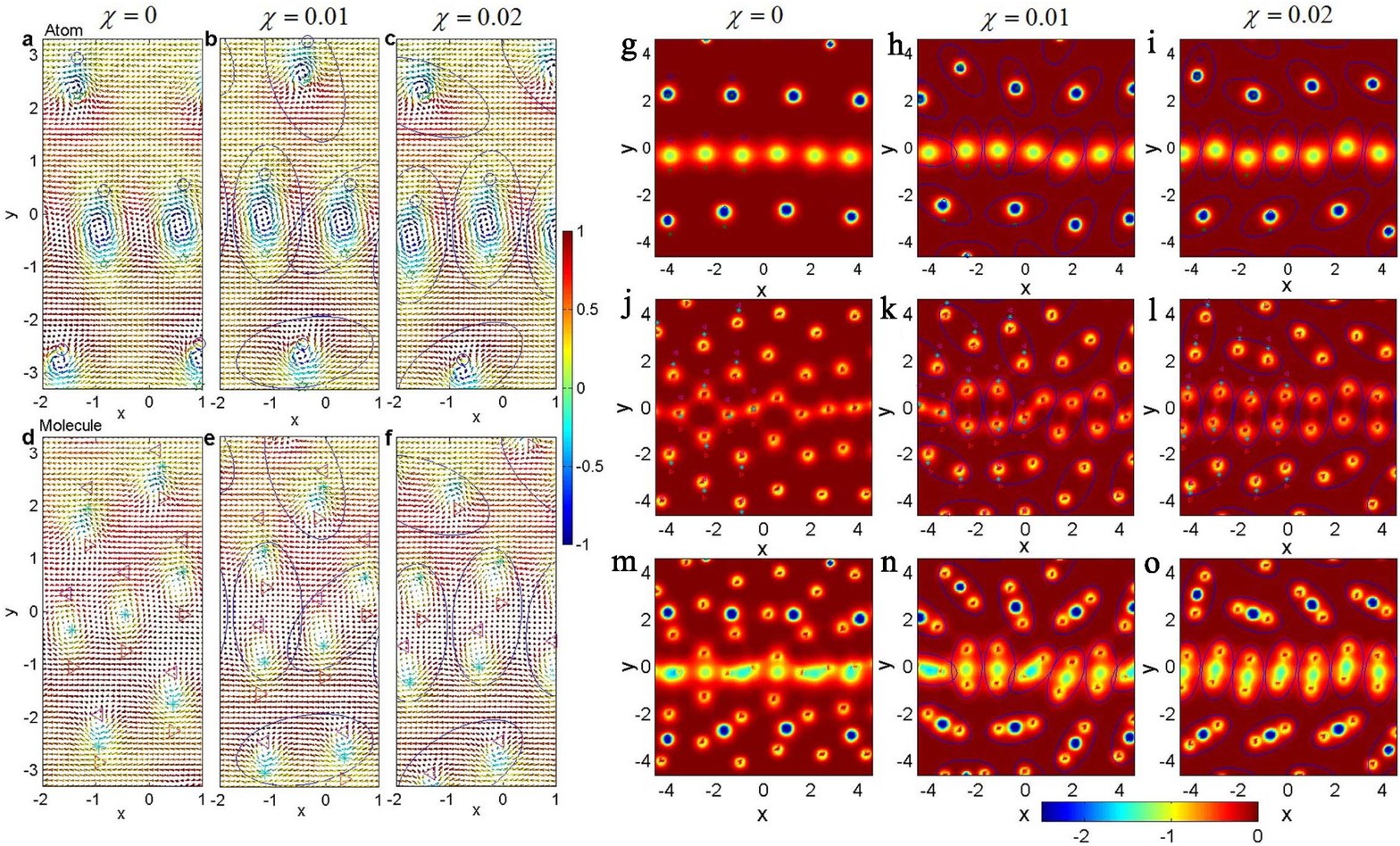,width=16cm} 
\end{center}
\label{fig5}
\end{figure}

\end{document}


\title{Supplementary Information for ``Carbon-dioxide-like Skyrmion controlled by spin-orbit coupling in atomic-molecular Bose-Einstein condensates''}

\author{Chao-Fei Liu$^{1,2}$, Gediminas Juzeli\={u}nas$^{3}$, and Wu-Ming Liu$^{1\star}$}
\maketitle

\begin{affiliations}
\item
Beijing National Laboratory for Condensed Matter Physics,
Institute of Physics, Chinese Academy of Sciences,
Beijing 100190, China

\item
School of Science, Jiangxi University of Science and Technology, Ganzhou 341000, China


\item
Institute of Theoretical Physics and Astronomy, Vilnius University, A. Go$\breve{s}$tauto 12, 01108 Vilnius, Lithuania


$^\star$e-mail: wliu@iphy.ac.cn
\end{affiliations}

%
%
%

In the paper, we have proposed and explored a way of creating a Rashba-Dresselhaus spin-orbit coupling (SOC) in an atomic-molecular Bose-Einstein condensates (BECs) by combining the spin dependent photoassociation and Raman coupling. Separately, the Raman coupling, which induces SOC \cite{Lind, Shuai-Chen2, Williams, Hamner, ChenYong, Zhangchuanwei}, and photoassociation, which produces molecular BEC \cite{ Jinghui, Donley, Hamley, Kobayashi, Heinzen, Wynar, Woo}, are well established experimental techniques, but there was no previous theoretical or experimental attempt to combine them. Amongst other results, we find a new type of topological excitation, the carbon-dioxide-like Skyrmion, in which two half-Skyrmions of the molecular BECs couple with one Skyrmion of the atomic BECs to form a carbon-dioxide-like structure. Our results show that the SOC can control the formation and distribution of the carbon-dioxide-like Skyrmion. This kind of Skyrmion can be measured via the vortices structures using the time-of-flight absorption imaging technique in the real experiment.

In this Supplementary Information, we first show the effects of atom-molecule conversion on the SOC atomic-molecular BECs, including consideration of the densities and the positions of vortices.
Furthermore, we explore the effect of Raman coupling on the spin textures in the SOC atomic-molecular BECs.
We prove that the carbon-dioxide-like Skyrmion would not occur in the normal rotating spinor atomic-molecular BECs without referring to both the Raman coupling and SOC.
Additionally, we demonstrate that the transformation $(\textbf{S}^{'}_{x}, \textbf{S}^{'}_{y}, \textbf{S}^{'}_{z})=(\textbf{S}_{z}, \textbf{S}_{y}, -\textbf{S}_{x})$ does not change the topological charge. Finally, the solutions for characterizing the carbon-dioxide-like Skyrmion are provided.

\subsection{The effect of atom-molecule conversion on the vortices in the spin-orbit coupling atomic-molecular BECs.}

Now, we perform some experiments where the atom-molecule conversion is changed to be 0, 0.01 and 0.02. Figures S1a-S1c show the densities of the SOC atomic-molecular BECs which is obtained under the equilibrium state. The five components are pointed out by the title: $|\uparrow\rangle$, $|\downarrow\rangle$, $m_{F}=-1$, $m_{F}=0$ and $m_{F}=+1$, respectively. Similar to the experiment in Fig. 3, there are lots of vortices inducing. Figures S1d-S1l indicate the position of vortices.
Here, the first column only shows the position of vortices formed by the atomic BECs (i.e., the spin-up and down components). The second column shows those formed by the molecular BECs (i.e., $m_{F}=0, \pm1$ components). The third column indicates all the positions of vortices. Note that we use three rectangles (the red, blue and black) to demonstrate the relationship between vortices and that among molecular vortices. In all of these plots, the three different rectangles are of the same sizes and the same positions, respectively.

It is easy to find that each vortex of spin-up component is just above a coupled vortex of spin-down component in Fig. S1d.
Off the $x$-axis, the distances between the two coupled vortices is very close (see the black rectangle), they become bigger in the blue rectangle, and they are the biggest in the red rectangle.
These phenomena also occur in Figs. S1g and S1i.
In Figs. S1e, S1h and S1k, the molecular vortices tend to form the three vortices structure where vortices of the $m_{F}=+1$, $m_{F}=0$, and $m_{F}=-1$ components couples and arranges in sequence. The black and blue rectangles in Figs. S1e and S1h do not catch some three vortices structures. But they can catch some three vortices structures in Fig. S1k.
Comparing the results in Figs. S1f, S1i and S1l, we can see clearly some vortex clusters and the vortices lattice (see the red rectangles) only in Fig. S1l. Therefore, the vortices position indicates that the creation of vortices cluster depends on the atom-molecule conversion.

The atom-molecule conversion enhances the coupling between atomic vortices and molecular ones. When this conversion is terminated (i.e., $\chi=0.0$), there is no coupling among atomic vortices and molecular vortices. In this situation, we obtain the regular distribution of atomic vortices in the black and blue rectangles and the relatively disordered molecular vortices in Fig. S1e. As the increase of the atom-molecule conversion, the molecular vortices have to unite with the atomic vortices and they are dragged into the black and blue rectangles. Thus, we can see the vortex clusters occurring in Fig. S1l. This effect also happens in the red rectangles and leads to the creation of vortices lattice.
Therefore, we can conclude that the atom-molecule conversion plays a role in coupling
the atomic vortices and the molecular vortices to form the vortex clusters and the vortices lattices.

In real experiment, the photoassociation process can cause the atom-molecule conversion \cite{ Jinghui, Donley, Hamley, Kobayashi, Heinzen, Wynar, Woo}. It is implies that the the photoassociation process are terminated if we choose the coefficient of atom-molecule conversion to be 0. In that case, it is impossible to obtain any molecular BECs as well as the corresponding topological excitations in them.
In this study, our method of damped projected Gross-Pitaevskii equations \cite{Rooney} will behavior just likes to catch the particles with a shallow trap and exchange particles with an external thermal reservoir. This process is kept on until the system reaches the equilibrium state.
So the molecular BECs is not completely coming from the trapped atomic BECs via the atom-molecule conversion.
However, our method can well demonstrate the effect of atom-molecule conversion on the vortices and Skyrmions in the SOC atomic-molecular BECs, because the external thermal reservoir only provides the particles for the system but not brings in any other effects.
Thus, the results of $\chi=0$ can be viewed as an experiment where one firstly have gotten a mixed atomic-molecular BECs, then the lasers for the spin-dependent photoassociation are turned off and the atomic-molecular BECs are further cooled to the ground state.
Therefore, the results are credible in Figs. S1a, S1d, S1e, S1f and in Figs. 5a, 5d, 5g, 5j, 5m.

\subsection{The effect of Raman coupling on the carbon-dioxide-like Skyrmion in the spin-orbit coupling atomic-molecular BECs.}

To clarify why the carbon-dioxide-like Skyrmion can be induced in the atomic-molecular BECs, we further investigate the effect of Raman coupling $\Omega_{A}$ on the vortices pattern and spin textures.
In Figs. S2a-S2c, we select the strength of the Raman coupling with the value of 0, $0.2\omega$ and $0.6\omega$, respectively.
Comparing these results, the phase separation between the two components of the atomic BECs is the most obvious
when the strength of the Raman coupling decreases to be 0.
The reason is that Raman coupling can make the spin-up and spin-down atoms convert with each other.
When we terminate this term ($\Omega_{A}=0$), the phase separation between two components of atomic BECs becomes the most obvious. Thus, there is little overlapping of densities between the two components of atomic BECs.
Similarly, the phase separation also happens in the molecular BECs, where the $m_{F}=0$ component is in the middle.
The corresponding vortices positions are shown in Figs. S2d-S2f. When the strength of the Raman coupling is 0 (see Fig. S2d), the carbon-dioxide vortices [the black rectangle and blue rectangle show two examples] are induced, and the regular structure of vortices lattices near the $x$-axis disappears [i.e., there is no vortices in the red rectangle]. When the Raman coupling increases to be $0.2\omega$ (see Fig. S2e), the red rectangle indicates that the regular structure of vortices lattices occurs near the $x$-axis. At the same time, vortices begin to form some clusters as shown by the black and blue rectangles.
Note that not all the vortices form the clusters.
When the Raman coupling is up to be $0.6\omega$ (see Fig. S2f), the vortices form lattices in the red rectangle and the formation of vortex clusters becomes more clearly than that in Fig. S2e.
Therefore, Raman coupling induces the coupling between atomic vortices of spin-up and spin-down components. Certainly, it also induces the coupling among the molecular vortices of the $m_{F}=0, \pm1$ components.

Raman coupling plays an important role in inducing the carbon-dioxide-like Skyrmion in atomic-molecular BECs.
Figures S3a-S3c (Figures S3d-S3f) show the corresponding spin textures of atomic BECs (molecular BECs) in Figs. S2a-S2c, respectively.
The spin textures are obtained under the transformation $(\textbf{S}^{'}_{x}, \textbf{S}^{'}_{y}, \textbf{S}^{'}_{z})=(\textbf{S}_{z}, \textbf{S}_{y}, -\textbf{S}_{x})$.
We use the blue ellipses to mark out the carbon-dioxide-like Skyrmions.
In Figs. S3a and S3d, there are circular structures of spin textures between the red lines. However, we would not like to view them as the effective Skyrmion structures.
Figures S4a and S4d show the topological charge densities of the spin textures in atomic BECs and those in molecular BECs.
The topological charge density in Fig. S4a shows that the Skyrmion structure in atomic BECs off $x$-axis is not very clear, though the carbon dioxide vortices exist. The main reason is that it is difficult for atomic BEC of spin up (down) components filling in the atomic vortices of spin down (up) component in the absence of Raman coupling.
While the spin textures along $x$-axis come from the domain wall of the atomic BECs. The topological charge density in Fig. S4a demonstrates a channel with a negligible change along the $x$-axis.
In fact, there is little vortices structure but the domain wall.
Here, we would not like to discuss this special structure.
Analogous cases also occur in Figs. S4d and S4g.
Certainly, the half-Skyrmions occur in Fig. S4d off $x$-axis. Figure S4g further proves that the structure of the carbon-dioxide-like Skyrmions is not very clear.
All these results are caused by the absence of Raman coupling. It is difficult for the components of atomic BECs (molecular BECs) to fill with each other. Thus, normal Skyrmions are restrained though some atomic vortices are induced. Naturally, it is hard to obtain the clear carbon-dioxide-like Skyrmions.

Raman coupling improves the creation of carbon-dioxide-like Skyrmions. In Fig. S3b, the structure of the Skyrmion (see the blue ellipses) is much more clear than that in Fig. S3a. We also use the red lines to indicate the region along the $x$-axis. The topological charge densities in Fig. S4b show that there are Skyrmions occurring along between the two red lines. However, at the same region we can not find the singular points in Fig. S4e. Thus, there is no complete half-Skyrmion induced along the $x$-axis. When the strength of Raman coupling increases to $0.6\omega$ (see Figs. S3c, S3f, S4c, S4f and S4i), there are complete half-Skyrmions whose topological charge densities have a singular point along the $x$-axis. Therefore, we can use the blue ellipses to mark out all the structures as the effective carbon-dioxide-like Skyrmions.
These results prove that the Raman coupling enhances the formation of carbon-dioxide-like Skyrmions.

\subsection{The ground state of the rotating atomic-molecular BECs without spin-orbit coupling.}

We now discuss the ground state of the rotating atomic-molecular BECs without SOC.
In our experimental setup [see Fig. 1 in the paper], the SOC originates from the counter propagating Raman lasers.
If we turn off the Raman lasers, the experimental setup would only induce the normal spinor atomic-molecular BECs.
Our purpose of this section is to prove that it is impossible to induce the carbon-dioxide-like Skyrmion in the rotating atomic-molecular BECs without SOC, i.e., without the counter propagating Raman lasers.
Differed from the case in Fig. 3a and Fig. S1a, both SOC and the Raman coupling are terminated.
Figures S5a and S5b show the densities distribution and the corresponding phases, respectively.
The vortices positions are indicated in Figs. S5c-S5e.
The spin textures show that there is no half-Skyrmions in Figs. S5f and S5g.
All the excitation of the spin textures are normal Skyrmions. In addition, the spin textures in atomic BECs are very similar to that in molecular BECs. The topological charge densities in Figs. S5h and S5i further prove that there is no half-Skyrmions. Meanwhile, the sum of the topological charge density does not show that there are any coupling between the spin textures of atomic BECs and molecular ones.
Therefore, the carbon-dioxide-like Skyrmion does not occur if the Raman lasers which induce SOC are not considered.

\subsection{Topological charge $Q$ is unchanged under a transformation $(\textbf{S}^{'}_{x}, \textbf{S}^{'}_{y}, \textbf{S}^{'}_{z})=(\textbf{S}_{z}, \textbf{S}_{y}, -\textbf{S}_{x})$.}

We now demonstrate that the transformation $(\textbf{S}^{'}_{x}, \textbf{S}^{'}_{y}, \textbf{S}^{'}_{z})=(\textbf{S}_{z}, \textbf{S}_{y}, -\textbf{S}_{x})$ does not change the value of the topological charge $Q$. It is well known that the topological charge density $q$ is defined as
\begin{equation}
q=\frac{1}{4\pi} \textbf{s}\cdot(\frac{\partial \textbf{s}}{\partial x}\times\frac{\partial \textbf{s}}{\partial y}),\label{s1}
\end{equation}
where $\textbf{s}=\textbf{S}/|\textbf{S}|$, and $Q=\int\int q dxdy$. For clarity, we use $q(\textbf{s}_{x}, \textbf{s}_{y}, \textbf{s}_{z})$ to describe the normal expression of the topological charge $q$ of Eq. (\ref{s1}), and $q(\textbf{s}_{z}, \textbf{s}_{y}, -\textbf{s}_{x})$ to denote the topological charge under the transformation $(\textbf{S}^{'}_{x}, \textbf{S}^{'}_{y}, \textbf{S}^{'}_{z})=(\textbf{S}_{z}, \textbf{S}_{y}, -\textbf{S}_{x})$.

We can expand the expression $q(\textbf{s}_{x}, \textbf{s}_{y}, \textbf{s}_{z})$ to be the form
\begin{eqnarray}
q(\textbf{s}_{x}, \textbf{s}_{y}, \textbf{s}_{z})=\frac{1}{4\pi}  \left|
  \begin{array}{ccc}
    \textbf{s}_{x} & \textbf{s}_{y} & \textbf{s}_{z} \\
    \frac{\partial}{\partial x} \textbf{s}_{x} & \frac{\partial}{\partial x} \textbf{s}_{y} & \frac{\partial}{\partial x} \textbf{s}_{z} \\
    \frac{\partial}{\partial y} \textbf{s}_{x} & \frac{\partial}{\partial y} \textbf{s}_{y} & \frac{\partial}{\partial y} \textbf{s}_{z} \\
  \end{array}
\right| . \notag
\end{eqnarray}
Similarly, $q(\textbf{s}_{z}, \textbf{s}_{y}, -\textbf{s}_{x})$ can be written as
\begin{eqnarray}
q(\textbf{s}_{z}, \textbf{s}_{y}, -\textbf{s}_{x})=\frac{1}{4\pi} \left|
  \begin{array}{ccc}
    \textbf{s}_{z} & \textbf{s}_{y} & -\textbf{s}_{x} \\
    \frac{\partial}{\partial x} \textbf{s}_{z} & \frac{\partial}{\partial x} \textbf{s}_{y} & -\frac{\partial}{\partial x} \textbf{s}_{x} \\
    \frac{\partial}{\partial y} \textbf{s}_{z} & \frac{\partial}{\partial y} \textbf{s}_{y} & -\frac{\partial}{\partial y} \textbf{s}_{x} \\
  \end{array}
\right| \notag
\end{eqnarray}
\begin{eqnarray}
    =\frac{1}{4\pi} \left|
  \begin{array}{ccc}
    \textbf{s}_{x} & \textbf{s}_{y} & \textbf{s}_{z} \\
    \frac{\partial}{\partial x} \textbf{s}_{x} & \frac{\partial}{\partial x} \textbf{s}_{y} & \frac{\partial}{\partial x} \textbf{s}_{z} \\
    \frac{\partial}{\partial y} \textbf{s}_{x} & \frac{\partial}{\partial y} \textbf{s}_{y} & \frac{\partial}{\partial y} \textbf{s}_{z} \\
  \end{array}
\right| =q(\textbf{s}_{x}, \textbf{s}_{y}, \textbf{s}_{z}). \notag
\end{eqnarray}
Thus, the transformation $(\textbf{S}^{'}_{x}, \textbf{S}^{'}_{y}, \textbf{S}^{'}_{z})=(\textbf{S}_{z}, \textbf{S}_{y}, -\textbf{S}_{x})$ does not change the value of the topological charge density $q$. Certainly, the topological charge $Q$ also is not changed under the transformation.

\subsection{Solution of carbon-dioxide-like Skyrmion.}
Here, we illuminate the solution of the carbon-dioxide-like Skyrmion. For clarify, we take the results in Fig. 2 as an example to discuss the solutions detailedly.
We find that the Skyrmion of atomic BECs in Fig. 2 can be represented approximately by
\begin{eqnarray} \label{SS1}
S_{1x}=-\frac{(0.75x)^{2}+y^{2}-4\times0.2 ^{2} e^{-2^2\times[(0.75x)^{2}+y^{2}]/2}}{(0.75x)^{2}+y^{2}+4\times0.2 ^{2} e^{-2^2\times[(0.75x)^{2}+y^{2}]/2}},\notag\\
S_{1y}=\frac{4 \times0.2\times 0.75x  e^{-2^2\times[(0.75x)^{2}+y^{2}]/2}}{(0.75x)^{2}+y^{2}+4\times0.2 ^{2} e^{-2^2\times[(0.75x)^{2}+y^{2}]/2}},\notag \\
S_{1z}=\frac{-4\times0.2\times y  e^{-2^2\times[(0.75x)^{2}+y^{2}]/2}}{(0.75x)^{2}+y^{2}+4\times0.2 ^{2} e^{-2^2\times[(0.75x)^{2}+y^{2}]/2}}.\\ \notag
\end{eqnarray}
Figure S6a plots the structure of the spin texture with Eqs. (\ref{SS1}).
Although the structure is different from the normal circular Skyrmion, it can transfer to the circular one under the transformation $(\textbf{S}^{'}_{x}, \textbf{S}^{'}_{y}, \textbf{S}^{'}_{z})=(\textbf{S}_{z}, \textbf{S}_{y}, -\textbf{S}_{x})$ [see Fig. S6b].
The topological charge density in Fig. S6c show that this spin texture are nontrivial Skyrmion with $Q=-1$.
In addition, we find that all the properties of this Skyrmion is very similar to that in Fig. 2.
Thus, Eqs. (\ref{SS1}) can well describe the center Skyrmion in the carbon-dioxide-like Skyrmion.

A single half-Skyrmion of molecular BECs in Fig. 2 can be represented by
\begin{eqnarray} \label{s2}
\textbf{s}_{2y}&=&\frac{ x}{\sqrt{x^{2}+ y^{2}}}\exp[-2.5^{2} (x^{2}+y^{2})/2], \notag\\
\textbf{s}_{2z}&=&-\frac{y}{\sqrt{x^{2}+y^{2}}}\exp[-2.5^2(x^{2}+y^{2})/2], \notag \\
\textbf{s}_{2x}&=&-\sqrt{1-\textbf{s}_{2y}^{2}-\textbf{s}_{2z}^{2}}.
\end{eqnarray}
Clearly, the two vectors $\textbf{s}_{2y}$, $\textbf{s}_{2z}$ can vary from $-1$ to $1$, but the vector $\textbf{s}_{2x}$ varies from $-1$ to $0$. This means that the spin vector can cover half of a unit sphere.
By calculating the topological charge $Q=\frac{1}{4\pi}\int\int \textbf{s}\cdot(\frac{\partial \textbf{s}}{\partial x}\times\frac{\partial \textbf{s}}{\partial y})dxdy$, where $\textbf{s}$ comes from Eqs. (\ref{s2}), we find the topological charge approaches $Q=-0.5$.

The two half-Skyrmion of the carbon-dioxide-like Skyrmion in Fig. 2 can be approximately written as
\begin{eqnarray} \label{s3}
\textbf{s}_{3y}&=&\frac{ (x-0.7)}{\sqrt{(x-0.7)^{2}+ y^{2}}}\exp[-2.5^{2} ((x-0.7)^{2}+y^{2})/2] \notag\\
&+&\frac{ (x+0.7)}{\sqrt{(x+0.7)^{2}+ y^{2}}}\exp[-2.5^{2} ((x+0.7)^{2}+y^{2})/2],\notag \\
\textbf{s}_{3z}&=&-\frac{y}{\sqrt{(x-0.7)^{2}+y^{2}}}\exp[-2.5^2((x-0.7)^{2}+y^{2})/2] \notag \\
&-&\frac{y}{\sqrt{(x+0.7)^{2}+y^{2}}}\exp[-2.5^2((x+0.7)^{2}+y^{2})/2], \notag\\
\textbf{s}_{3x}&=&-\sqrt{1-\textbf{s}_{3y}^{2}-\textbf{s}_{3z}^{2}}.
\end{eqnarray}
The structure of spin texture with Eqs. (\ref{s3}) is shown in Fig. S6d. If we perform the transformation:
$(\textbf{S}^{'}_{x}, \textbf{S}^{'}_{y}, \textbf{S}^{'}_{z})=(\textbf{S}_{z}, \textbf{S}_{y}, -\textbf{S}_{x})$,
we can obtain two clear circular structure [see Fig. S6e].
Comparing it with Fig. 2f, the formation of the two spin texture is very similar.
This indicates the half-Skyrmion can be well described by Eqs. (\ref{s3}) in our simulation.
Furthermore, the topological charge density in Fig. S6f is very similar to that in Fig. 2j.
Therefore, Eqs. (\ref{s3}) can well describe the two half-Skyrmions in the carbon-dioxide-like Skyrmion.

\newpage

\textbf{Supplementary Figure S1 The effect of atom-molecule conversion on the vortices in the SOC atomic-molecular BECs.}
The rotation frequency is $\Omega_{rotation}=0.5\omega$, where $\omega=200\times2\pi$Hz.
We set the parameters $\lambda=1$, $\Omega_{A}=0.8$, $g_{\uparrow,\uparrow}=g_{\uparrow,\downarrow}=g_{A}$ with the scattering length $a_{A}=101.8a_{B}$,
$g_{\downarrow,\downarrow}=0.95g_{A}$, $g_{n}=2g_{A}$, $g_{s}=0.03g_{A}$, $g_{AM}=0.5g_{A}$, $\varepsilon=0$ and $\mu_{m}=2\mu_{a}=3.5\mu_{m,0}=7\mu_{a,0}=42\hbar\omega$.
(\textbf{a}-\textbf{c}) show the densities of each component when the system reaches the equilibrium state. The components are pointed out by the title: $|\uparrow\rangle$, $|\downarrow\rangle$, $m_{F}=-1$, $m_{F}=0$ and $m_{F}=+1$, respectively. (\textbf{a}) $\chi=0.0$,
(\textbf{b}) $\chi=0.01$, and (\textbf{c}) $\chi=0.02$.
(\textbf{d}-\textbf{l}) The position of vortices. The title `atom', `molecule' and `total' indicate that the first, second and third column plot the position of vortices in atomic BECs, molecular ones and the both, respectively.
(\textbf{d}) The position of vortices formed by atomic BECs in (\textbf{a}). (\textbf{e}) The position of vortices formed by molecular BECs in (\textbf{a}).
(\textbf{f}) shows all vortices' position in (\textbf{a}).
The red, blue and black rectangles are used to mark out that atomic vortices tend to march with each other, and distribute parallel with the $x$ axis.
Similarly, (\textbf{g}-\textbf{i}) show the corresponding position of vortices in (\textbf{b}).
(\textbf{j}-\textbf{l}) show the corresponding position of vortices in (\textbf{c}).
The particles number of each component $(N_{\uparrow}, N_{\downarrow}, N_{-1}, N_{0}, N_{+1})$
are $(6.7\times10^{3}, 7.7\times10^{3}, 8.0\times10^{3}, 3.8\times10^{3}, 8.1\times10^{3})$ in (\textbf{a}),
$(6.9\times10^{3}, 7.8\times10^{3}, 8.0\times10^{3}, 3.8\times10^{3}, 8.1\times10^{3})$ in (\textbf{b}),
and $(7.0\times10^{3}, 7.8\times10^{3}, 7.9\times10^{3}, 4.0\times10^{3}, 7.9\times10^{3})$ in (\textbf{c}).
The unit of length, strength of $\chi$, $\Omega_{A}$ and SOC are $0.76\mu m$, $\hbar\omega$, $\omega$, and $0.96\times10^{-3}m/s$, respectively.

\bigskip
\textbf{Supplementary Figure S2 The effect of Raman coupling on the vortices in the SOC atomic-molecular BECs.}
The  rotation frequency is $\Omega_{rotation}=0.5\omega$, where $\omega=200\times2\pi$Hz.
We set the parameters $\lambda=1$, $g_{\uparrow,\uparrow}=g_{\uparrow,\downarrow}=g_{A}$ with the scattering length $a_{A}=101.8a_{B}$,
$g_{\downarrow,\downarrow}=0.95g_{A}$, $g_{n}=2g_{A}$, $g_{s}=0.03g_{A}$, $g_{AM}=0.5g_{A}$, $\chi=0.08$, $\varepsilon=0$ and $\mu_{m}=2\mu_{a}=3.5\mu_{m,0}=7\mu_{a,0}=42\hbar\omega$.
All the results are obtained when the system reaches the equilibrium state. The five components are pointed out by the title: $|\uparrow\rangle$, $|\downarrow\rangle$, $m_{F}=-1$, $m_{F}=0$ and $m_{F}=+1$, respectively. (\textbf{a}-\textbf{c}) The effect of Raman coupling on the densities of the SOC atomic-molecular BECs. (\textbf{a}) $\Omega_{A}=0$, (\textbf{b}) $\Omega_{A}=0.2$, and (\textbf{c}) $\Omega_{A}=0.6$.
(\textbf{d}-\textbf{f}) The corresponding position of vortices in (\textbf{a}-\textbf{c}), respectively. The blue ellipses point out the regions of the carbon-dioxide-like Skyrmion. In addition, we mark the position of vortices in order to illuminate
the relationship between spin texture and position of vortices clearly. The meanings of the marks are the same as that in Figs. S2\textbf{d}-S2\textbf{f}.
The particles number of each component $(N_{\uparrow}, N_{\downarrow}, N_{-1}, N_{0}, N_{+1})$
are $(8.0\times10^{3}, 8.6\times10^{3}, 8.6\times10^{3}, 2.1\times10^{3}, 8.6\times10^{3})$, $(7.8\times10^{3}, 8.8\times10^{3}, 8.6\times10^{3}, 2.4\times10^{3}, 8.3\times10^{3})$, and $(8.0\times10^{3}, 8.6\times10^{3}, 8.1\times10^{3}, 3.4\times10^{3}, 8.1\times10^{3})$, respectively.
The unit of length, strength of $\chi$, $\Omega_{A}$ and SOC are $0.76\mu m$, $\hbar\omega$, $\omega$, and $0.96\times10^{-3}m/s$, respectively.

\bigskip
\textbf{Supplementary Figure S3 The effect of Raman coupling on the spin textures in the SOC atomic-molecular BECs.} The spin textures are obtained under the transformation $(\textbf{S}^{'}_{x}, \textbf{S}^{'}_{y}, \textbf{S}^{'}_{z})=(\textbf{S}_{z}, \textbf{S}_{y}, -\textbf{S}_{x})$. The color of each arrow indicates the magnitude of $S^{'}_{z}$. (\textbf{a}, \textbf{d}) The corresponding spin textures of atomic BECs and molecular BECs in Fig. S2a. (\textbf{b}, \textbf{e}) The spin textures of atomic BECs and molecular BECs in Fig. S2b. (\textbf{c}, \textbf{f}) The spin textures of atomic BECs and molecular BECs in Fig. S2c.
The blue ellipses point out the regions of the carbon-dioxide-like Skyrmions. The position of vortices are marked out with the same symbols as that in Figs. S2\textbf{d}-S2\textbf{f}.
The unit of length and SOC are $0.76\mu m$ and $0.96\times10^{-3}m/s$, respectively.

\bigskip

\textbf{Supplementary Figure S4 The effect of Raman coupling on the topological charge densities in the SOC atomic-molecular BECs.}
(\textbf{a}-\textbf{f}) The corresponding topological charge densities of the spin texture in Figs. S3\textbf{a}-S3\textbf{f}, respectively.
(\textbf{g}) The total topological charge densities of (\textbf{a}) and (\textbf{d}). (\textbf{h}) The total topological charge density of (\textbf{b}) and (\textbf{e}). (\textbf{i}) The total topological charge density of (\textbf{c}) and (\textbf{f}).
The blue ellipses point out the regions of the carbon-dioxide-like Skyrmions.
In the region of $x<0$, the position of vortices are marked out. The meanings of the marks are the same as that in Figs. S2\textbf{d}-S2\textbf{f}.
The unit of length and SOC are $0.76\mu m$ and $0.96\times10^{-3}m/s$, respectively.

\bigskip
\textbf{Supplementary Figure S5 The ground state of rotating atomic-molecular BECs without SOC.}
The rotation frequency is $\Omega_{rotation}=0.5\omega$, where $\omega=200\times2\pi$Hz.
We set the parameters $\lambda=0$, $\Omega_{A}=0$, $g_{\uparrow,\uparrow}=g_{\uparrow,\downarrow}=g_{A}$ with the scattering length $a_{A}=101.8a_{B}$, $g_{\downarrow,\downarrow}=0.95g_{A}$, $g_{AM}=0.5g_{A}$, $g_{n}=2g_{A}$, $g_{s}=0.03g_{A}$, $\chi=0.08$, $\varepsilon=0$ and $\mu_{m}=2\mu_{a}=3.5\mu_{m,0}=7\mu_{a,0}=28\hbar\omega$.
(\textbf{a}) Densities of each components under the equilibrium state. Note that the components are pointed out by the title: $|\uparrow\rangle$, $|\downarrow\rangle$, $m_{F}=-1$, $m_{F}=0$ and $m_{F}=+1$, respectively.
(\textbf{b}) The corresponding phases of each components in (\textbf{a}).
(\textbf{c}-\textbf{e}) The position of vortices.
(\textbf{c}) The position of vortices formed by atomic BECs. (\textbf{d}) The position of vortices formed by molecular BECs.
(\textbf{e}) shows all vortices' position.
(\textbf{f}) The spin textures of atomic BECs under the transformation $(\textbf{S}^{'}_{x}, \textbf{S}^{'}_{y}, \textbf{S}^{'}_{z})=(\textbf{S}_{z}, \textbf{S}_{y}, -\textbf{S}_{x})$.
(\textbf{g}) The corresponding spin textures of  molecular BECs. The color of each arrow indicates the magnitude of $S^{'}_{z}$.
(\textbf{h}) The topological charge densities of the spin texture in Figs. S5\textbf{f}.
(\textbf{i}) The topological charge densities of the spin texture in Figs. S5\textbf{g}.
(\textbf{j}) The total topological charge densities of (\textbf{h}) and (\textbf{i}).
The particles number $(N_{\uparrow}, N_{\downarrow}, N_{-1}, N_{0}, N_{+1})$
is $(2.72\times10^{3}, 3.35\times10^{3}, 2.48\times10^{3}, 2.80\times10^{3}, 2.23\times10^{3})$.
The unit of length, strength of $\chi$, $\Omega_{A}$ and SOC are $0.76\mu m$, $\hbar\omega$, $\omega$, and $0.96\times10^{-3}m/s$, respectively.

\bigskip
\textbf{Supplementary Figure S6 Solution of carbon-dioxide-like Skyrmion.}
(\textbf{a}) Structure of a Skyrmion with Eqs. (\ref{SS1}). The color of each arrow indicates the magnitude of $S_{1z}$.
(\textbf{b}) Structure of a Skyrmion with Eqs. (\ref{SS1}) under the transformation $(\textbf{S}^{'}_{x}, \textbf{S}^{'}_{y}, \textbf{S}^{'}_{z})=(\textbf{S}_{z}, \textbf{S}_{y}, -\textbf{S}_{x})$. The color of each arrow indicates the magnitude of $-S_{1x}$.
(\textbf{c}) The topological charge density $q$ with Eqs. (\ref{SS1}).
(\textbf{d}) Structure of two half-Skyrmions with Eqs. (\ref{s3}). The color of each arrow indicates the magnitude of $s_{3z}$.
(\textbf{e}) Structure of two half-Skyrmions with Eqs. (\ref{s3}) under the transformation $(\textbf{S}^{'}_{x}, \textbf{S}^{'}_{y}, \textbf{S}^{'}_{z})=(\textbf{S}_{z}, \textbf{S}_{y}, -\textbf{S}_{x})$. The color of each arrow indicates the magnitude of $-s_{3x}$.
(\textbf{f}) The topological charge density $q$ with Eqs. (\ref{s3}).

\newpage

\begin{figure}
\begin{center}
\epsfig{file=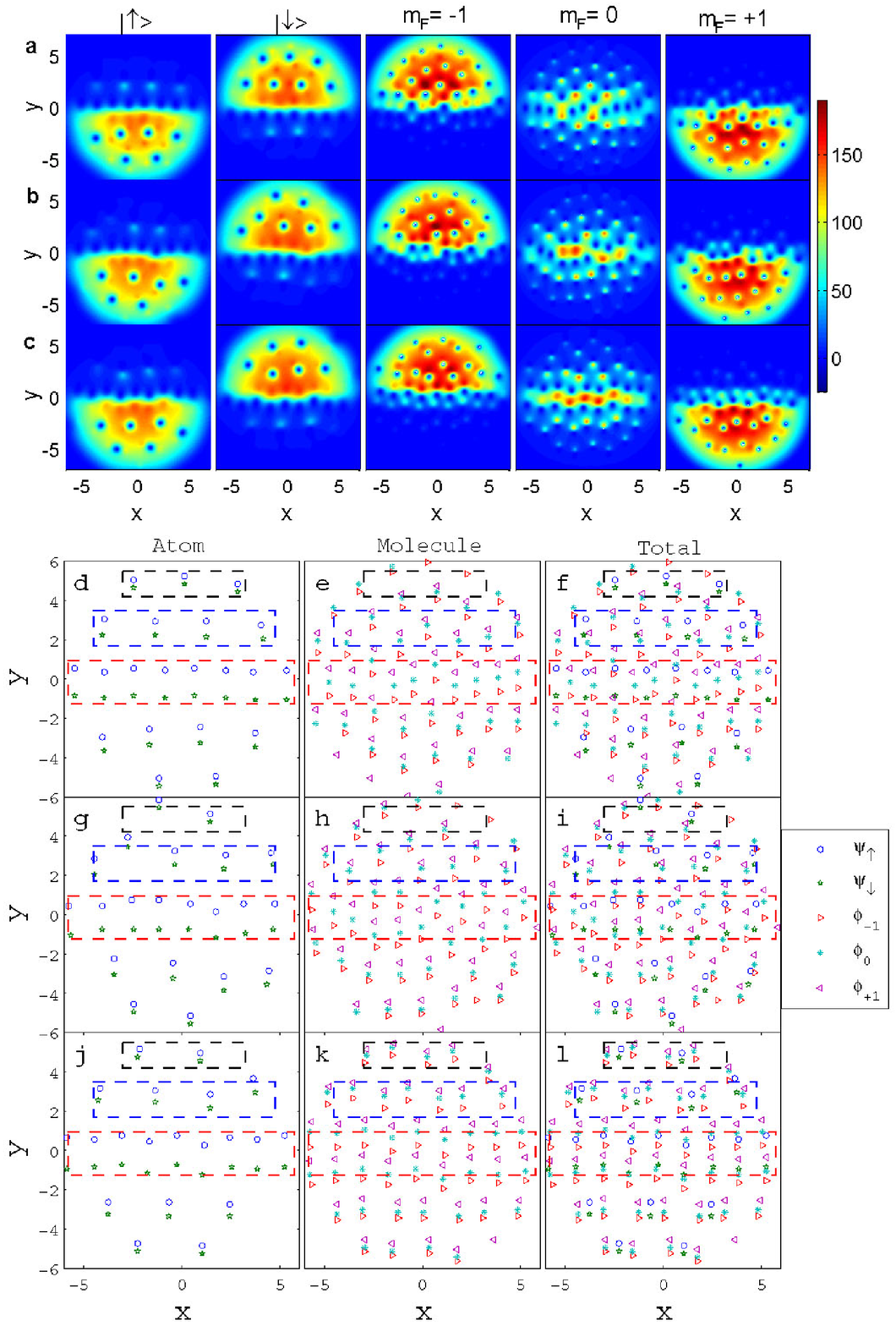,width=13cm} 
\end{center}
\label{fi}
\end{figure}

\begin{figure}
\begin{center}
\epsfig{file=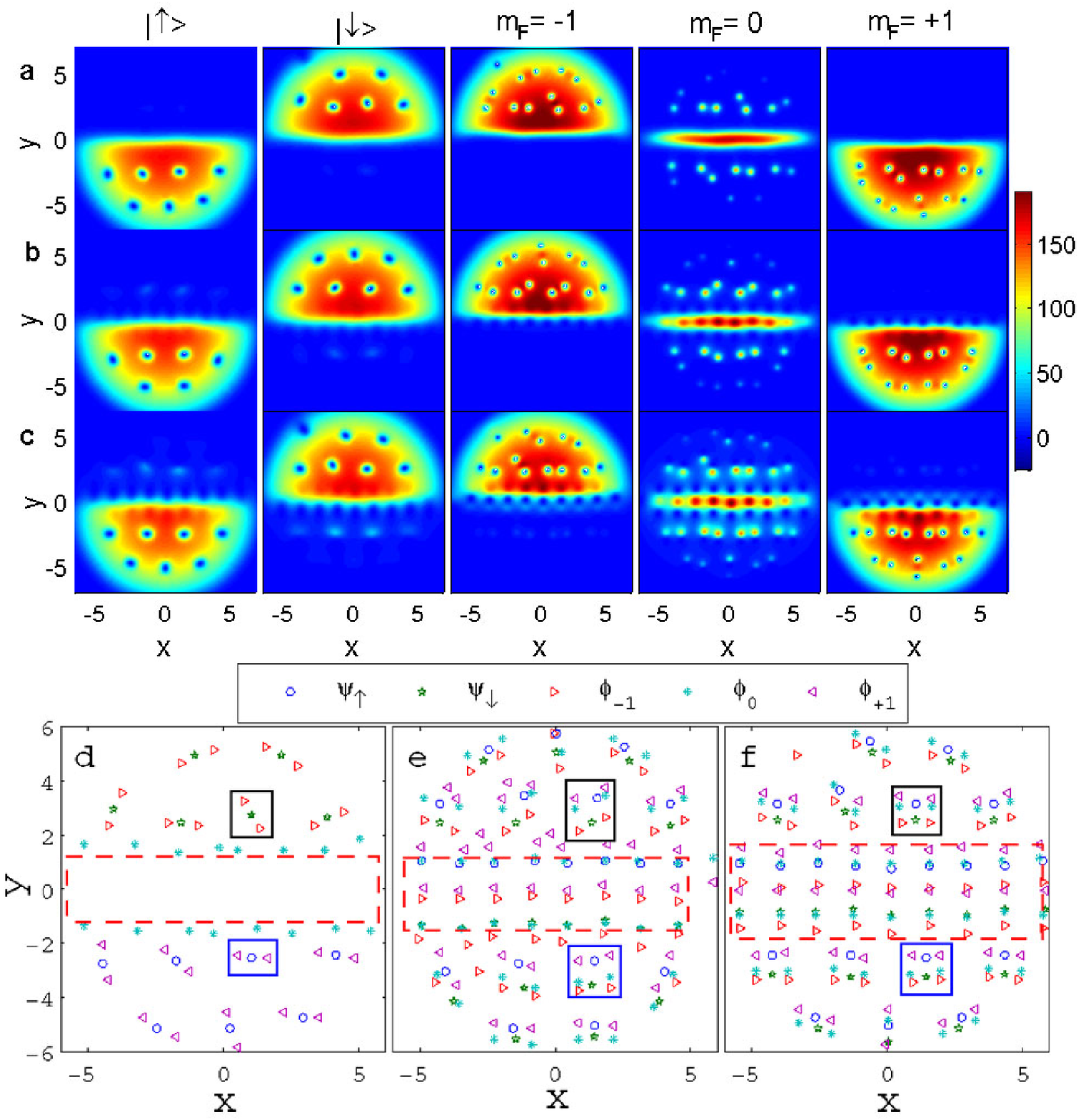,width=13cm} 
\end{center}
\label{fig:FiniteT}
\end{figure}


\begin{figure}
\begin{center}
\epsfig{file=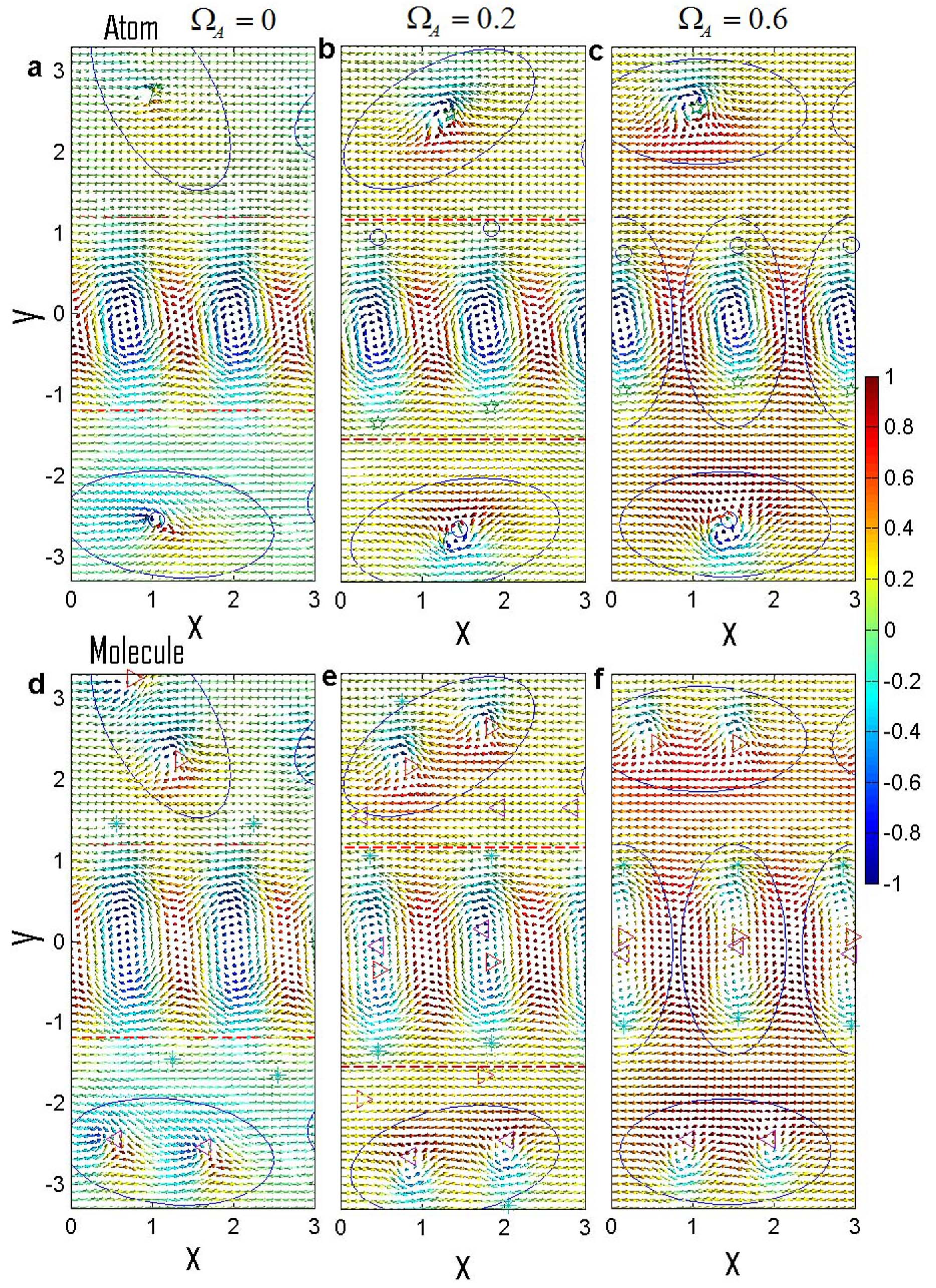,width=12.5cm} 
\end{center}
\label{fig:FiniteT}
\end{figure}

\begin{figure}
\begin{center}
\epsfig{file=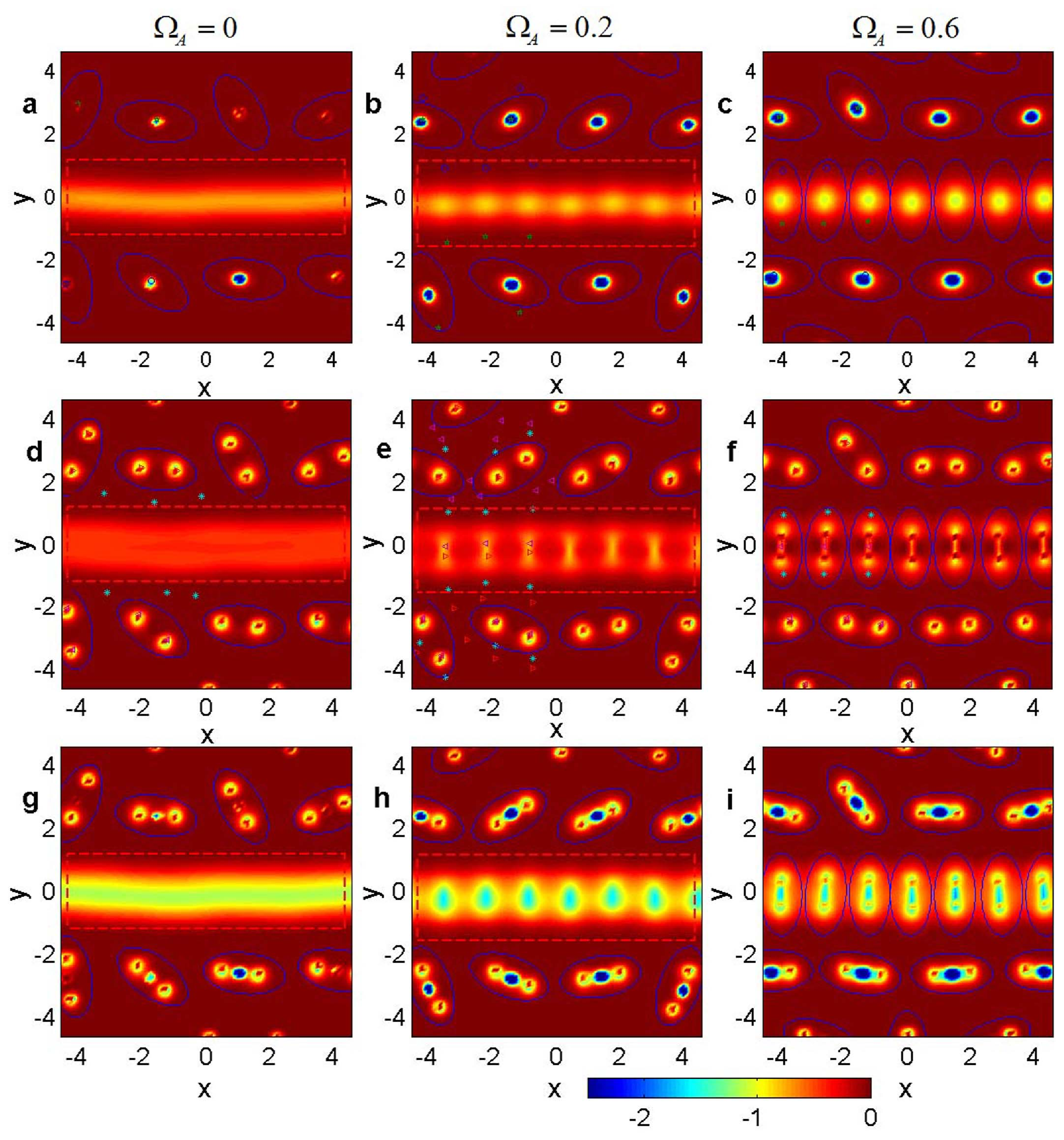,width=12.5cm} 
\end{center}
\label{fig:FiniteT}
\end{figure}

\begin{figure}
\begin{center}
\epsfig{file=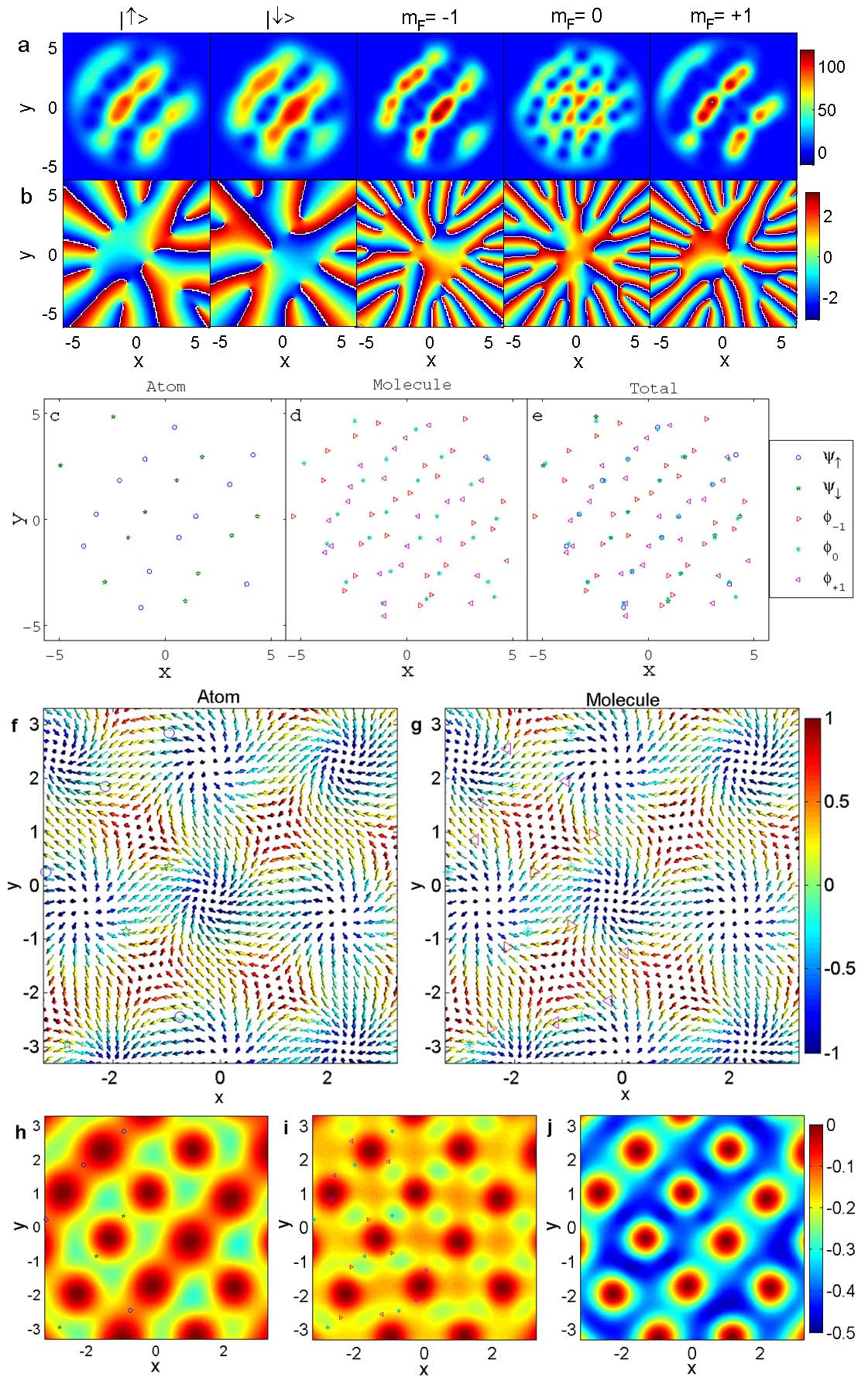,width=12.5cm} 
\end{center}
\label{fig:FiniteT}
\end{figure}

\begin{figure}
\begin{center}
\epsfig{file=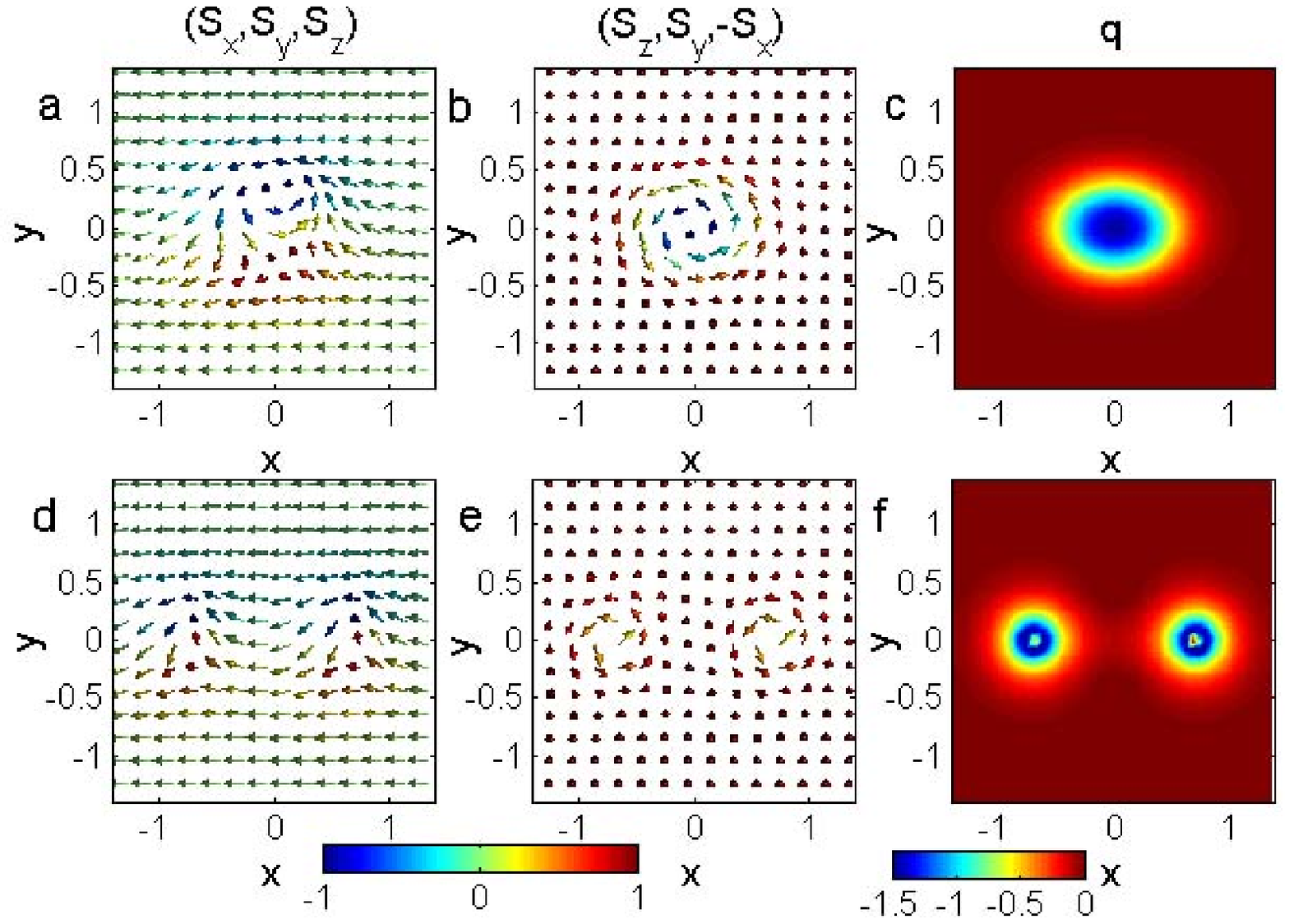,width=12.5cm} 
\end{center}
\label{fig:FiniteT}
\end{figure}